\documentclass[fleqn,usenatbib]{mnras}

\usepackage{newtxtext,newtxmath}

\usepackage[T1]{fontenc}

\DeclareRobustCommand{\VAN}[3]{#2}
\let\VANthebibliography\thebibliography
\def\thebibliography{\DeclareRobustCommand{\VAN}[3]{##3}\VANthebibliography}

\usepackage{graphicx}	
\usepackage{amsmath}	

\title[MOMO V: OJ 287 from radio to $\gamma$-rays]{MOMO V. Effelsberg, Swift and Fermi study of the blazar and supermassive binary black hole candidate OJ 287 in a period of high activity}

\author[S. Komossa et al.]{
S. Komossa$^{1}$\thanks{E-mail: astrokomossa@gmx.de (SK)},
D. Grupe$^{2}$, 
A. Kraus$^{1}$, A. Gonzalez$^{3}$, L.C. Gallo$^{3}$, M.J. Valtonen$^{4,5}$, 
\newauthor
S. Laine$^6$, T.P. Krichbaum$^{1}$, M.A. Gurwell$^7$, J.L. G\'omez$^{8}$,  
S. Ciprini$^{9,10}$, 
\newauthor
I. Myserlis$^{11,1}$, U. Bach$^{1}$
\\
$^{1}$Max-Planck-Institut f\"ur Radioastronomie, Auf dem H{\"u}gel 69, 53121 Bonn, Germany\\
$^{2}$Dept. of Physics, Earth Science, and Space System Engineering, Morehead State University, 235 Martindale Dr, Morehead, KY 40351, USA\\
$^{3}$Department of Astronomy and Physics, Saint Mary’s University, 923 Robie Street, Halifax, NS, B3H 3C3, Canada\\
$^{4}$ Finnish Centre for Astronomy with ESO, University of Turku, FI-20014, Turku, Finland \\
$^{5}$Department of Physics and Astronomy, University of Turku, FI-20014, Turku, Finland \\
$^{6}$IPAC, Mail Code 314-6, Caltech, 1200 E. California Blvd., Pasadena, CA 91125, USA \\
$^{7}$Center for Astrophysics | Harvard \& Smithsonian, Cambridge, MA 02138, USA \\
$^{8}$Instituto de Astrofísica de Andalucía-CSIC, Glorieta de la Astronomía s/n, E-18008 Granada, Spain \\
$^9${Istituto Nazionale di Fisica Nucleare (INFN) Sezione di Roma Tor Vergata, Via della Ricerca Scientifica 1, 00133, Roma, Italy} \\
$^{10}$ASI Space Science Data Center (SSDC), Via del Politecnico, 00133, Roma, Italy \\
$^{11}$Institut de Radioastronomie Millim\'etrique, Avenida Divina Pastora 7, Local 20, 18012, Granada, Spain
 } 

\date{Accepted 2022 March 18. Received 2022 March 1; in original form 2022 January 20}

\pubyear{2022}

\begin{document}
\label{firstpage}
\pagerange{\pageref{firstpage}--\pageref{lastpage}}
\maketitle

\begin{abstract}
We report results from our ongoing project MOMO (Multiwavelength Observations and Modelling of OJ 287).
In this latest publication of a sequence, we combine our Swift UVOT--XRT and Effelsberg radio data (2.6--44 GHz) between 2019 and 2022.04 with public SMA data and $\gamma$-ray data from the Fermi satellite. 
The observational epoch covers OJ 287 in a high state of activity from radio to X-rays. The epoch also covers two major events predicted by the binary supermassive black hole (SMBH) model of OJ 287. 
Spectral and timing analyses clearly establish: a 
new UV--optical minimum state in 2021 December at an epoch where
the secondary SMBH is predicted to cross the disk surrounding the primary SMBH;  
an overall low level of $\gamma$-ray activity
in comparison to pre-2017 epochs; the presence of a remarkable, long-lasting UV--optical flare event of intermediate amplitude in 2020--2021; 
a high level of activity in the radio band with multiple flares; and particularly a bright, ongoing radio flare peaking in 2021 November that may be associated with a $\gamma$-ray flare, the strongest in 6 years. 
Several explanations for the UV--optical minimum state are explored, including the possibility that a secondary SMBH launches a temporary jet, but the observations are best explained by variability associated with the main jet. 
\end{abstract}





\section{Introduction}

Supermassive binary black holes (SMBBHs) form in the course of galaxy mergers. They are expected to be the loudest sources of low-frequency gravitational waves (GWs) in the Universe \citep{Sesana2021} and they are a key component in our understanding of galaxy evolution \citep{Volonteri2003, Komossa2016a}. The search for, and analysis of, SMBBHs in all stages of their evolution has therefore evolved 
into an important field in extragalactic astrophysics. The most evolved binaries are well beyond the ``final parsec'' in their orbital evolution \citep{Colpi2014} where GW emission contributes to their orbital shrinkage. They have so far remained spatially unresolved with available imaging techniques, and indirect methods are used for their search and identification.  

Blazars are particularly
suitable for searching sub-parsec binary systems, because of periodicity imprints on light curves due to beaming effects 
and on jet structures \citep{Begelman1980}, and many candidate binaries among blazars have been identified in recent years \citep[e.g.,][]{Sillanpaa1988, Graham2015}. Different mechanisms to explain semi-periodicities have been considered in blazars: apparent changes in the observed luminosity because of
periodically varying Doppler boosting due to a precessing jet or due to an orbiting secondary SMBH, or true changes in the intrinsic luminosity due to disk impacts  (tilted orbit of the secondary) or due to stream-feeding from circumbinary disks in systems without inner accretion disk (in-plane orbit of the secondary) \citep[see][for reviews]{deRosa2019, Komossa2021b}.

These signatures do not always uniquely imply a binary. For instance, disk precession around a single SMBH can lead to semi-periodicity in light curves as well \citep{Liska2018}. 
Periodic jet structures can also arise around single SMBHs through helical magnetic fields and MHD instabilities \citep{Rieger2004, Lister2013}. Therefore, multiwavelength observations and long-term monitoring efforts are required to distinguish between different possible causes of semi-periodicities \citep{Valtonen2022}.  

Independent of the presence of any binary, blazars are excellent laboratories for understanding disk-jet physics \citep{Blandford2019}.  
Hence, detailed observations of individual, bright, nearby, and active systems are extremely important in furthering our understanding of
the geometry and emission mechanisms of blazars in general and SMBBHs in particular. 
 
OJ 287 is such a nearby, bright blazar (redshift $z$=0.306, R.A.: 08$^{\rm h}$54$^{\rm m}$48.87$^{\rm s}$, decl: +20$^{\circ}$06$\arcmin$30.6$\arcsec$) and among the best candidates to date for hosting a compact SMBBH.
Its optical light curve was characterized by sharp and bright outbursts 
increasing flux more than tenfold, and lasting for months \citep{Sillanpaa1988}. Optical maxima are double-peaked \citep{Sillanpaa1996}. Based on the semi-periodic appearance of the bright maxima, several variants of SMBBH models were considered \citep[e.g.][]{Sillanpaa1988, Lehto1996,  Katz1997, Villata1998, Valtaoja2000,   
Liu2002, Qian2015, Britzen2018, Dey2018}. The best-explored scenario,  modelled in most detail and making predictions for future events, involves a binary with a massive primary SMBH of $1.8\times10^{10}$
M$_{\odot}$, and a secondary SMBH of $1.5\times10^8$ M$_{\odot}$
on an eccentric, precessing orbit \citep{Valtonen2016, Dey2018, Laine2020, Valtonen2022}. The double-peaks in the optical lightcurve are explained as the  
times when the secondary SMBH impacts the disk around the primary twice during its 12.06 yr orbit \citep[``impact flares" hereafter;][]{Lehto1996, Valtonen2019}. The main flares do not become visible immediately, but only after the impact-driven bubbles expanding from the impact point \citep{Ivanov1998} become optically thin. 

OJ 287 is also a good example of a  multi-messenger source.
In the binary scenario, GWs are expected to be detected directly with future pulsar timing arrays \citep[PTAs;][]{Yardley2010, Valtonen2022}.
OJ 287 is a bright and variable emitter across the whole electromagnetic spectrum. In the optical band, its brightest outbursts reached 12 mag, comparable to the brightness of one of the nearest quasars.
OJ 287 is detected at VHE energies ($E>100$ GeV) with VERITAS \citep{OBrien2017} and in the $\gamma$-ray regime with the Fermi \citep{Abdo2009, Ballet2020}.
In X-rays, the first detection was with Einstein \citep{Madejski1988}, then followed by observations with other major X-ray observatories 
(per observatory, presented first by \citet[][EXOSAT]{Sambruna1994}, 
\citet[][ROSAT]{Comastri1995}, 
\citet[][ASCA]{Idesawa1997},
\citet[][Swift]{Massaro2003}, \citet[][XMM-Newton]{Ciprini2007}, \citet[][BeppoSAX]{Massaro2008}, \citet[][Ginga]{Seta2009}, \citet[][Chandra]{Marscher2011}, \citet[][NuSTAR]{Komossa2020}, and  \citet[][AstroSat]{Prince2021}, respectively).  

Observations with XMM-Newton and the Neil Gehrels Swift observatory (Swift hereafter) over two decades established 
OJ 287 as one of the most spectrally variable blazars in the X-ray band \citep{Komossa2020, Komossa2021a, Komossa2021d}. XMM-Newton spectroscopy has firmly established the presence of both a highly variable synchrotron and inverse Compton (IC) component, offering a unique chance of observing {\em{both}} components in the same soft X-ray band; rare in blazars \citep{Komossa2020, Komossa2021a}.  
Outbursts in 2016/17 \citep{Komossa2017} and 2020 \citep{Komossa2020}, discovered in the course of dedicated Swift monitoring of OJ 287, are driven by supersoft synchrotron flares \citep{Komossa2020}. NuSTAR discovered a remarkable steep X-ray spectrum up to 70 keV \citep{Komossa2020}. OJ 287 hosts an extended X-ray jet detected with Chandra \citep{Marscher2011}. 
Structure function analyses revealed characteristic timescales of 4–39 days
depending on wave band and activity state of OJ 287 \citep{Komossa2021d}.   

OJ 287 is a strong and highly variable radio source and has been the target of past radio-monitoring efforts,  
including campaigns between 1972 and 1996  \citep{Valtaoja2000}, and
it was part of a multi-frequency radio monitoring campaign of a larger sample of Fermi AGN between 2007 and 2014
\citep{Fuhrmann2016, Hodgson2017}. 
OJ 287 was also subject of major concerted optical monitoring campaigns, including polarimetry at select epochs. These campaigns were essential in identifying the sharp double-peaks over decades and in developing the binary model \citep[e.g.,][and references therein]{Smith1987, Sillanpaa1988, Sillanpaa1996, 
Pursimo2000, Villforth2010, Valtonen2011, Pihajoki2013a, Zola2016, Dey2018, Valtonen2022}, and in tracing the short-time variability characteristics with the Kepler mission \citep{Wehrle2019}.    
However, these important campaigns still lacked the broad-band coverage that is necessary to distinguish between various different thermal and non-thermal emission processes from the jet, the accretion disk and the binary at all times. For instance, optical polarization can be variable in both scenarios; mixtures of thermal and non-thermal emission 
on the one hand, and pure non-thermal shocks in jets on the other hand. $\gamma$-ray and radio monitoring traces the non-thermal emission well but misses thermal components as well as the non-thermal spectral energy distribution (SED) around its peak. 
Further, the SMBBH model continues to make new predictions that can only be tested in future observations \citep[e.g.,][]{Valtonen2022}.  

Therefore, the program MOMO (Multiwavelength Observations and Monitoring of OJ 287) was initiated in late 2015 \citep[e.g.,][]{Komossa2021c}. The 100m radio telescope at Effelsberg and the Neil Gehlrels Swift observatory are at the heart of the project and are combined with deep follow-up multiwavelength spectroscopy and public $\gamma$-ray observations. Together they cover frequencies between 2 GHz and 100 GeV.
MOMO provides broad-band SEDs, light curves and spectra in all activity states of OJ 287. 

This paper is the latest in a sequence reporting about MOMO results.
Here, we present Effelsberg and Swift observations between 2019 January and 2021 December, covering several epochs of exceptional binary and/or jet activity.  
This paper is structured as follows.
In Sect. 2 we introduce the MOMO project; its key goals, its set-up and the main results obtained so far. In Sect. 3 we present the analysis and results from the Swift data. The Effelsberg multifrequency radio data and results are given in Sect. 4. The discussion is presented in Sect. 5 with focus on epochs of exceptional flux or spectral states, and epochs of predicted binary activity. Summary and conclusions are provided in Sect. 6. 
Throughout this paper we use a cosmology with
$H_{\rm 0}$=70 km\,s$^{-1}$\,Mpc$^{-1}$, $\Omega_{\rm M}$=0.3 and $\Omega_{\rm \Lambda}$=0.7. At the distance of OJ 287, this corresponds to a scale of 4.5 kpc/arcsec \citep{Wright2006}. 

\section{MOMO project} 

\subsection{Program description and previous results} 
The program MOMO consists of dedicated, dense, long-term flux and spectroscopic monitoring 
and deep, higher-sensitivity follow-up 
observations of the blazar OJ 287 at $>$13 frequencies from the radio to the high-energy band \citep[see][for an overview]{Komossa2021c}.  
It was initiated in late 2015.
In particular, we are using Swift to cover the optical and UV bands in six filters as well as the 0.3-10 keV X-ray band,  
and we are using the 100m Effelsberg radio telescope to acquire radio measurements between 2.6 and 44 GHz.
Public $\gamma$-ray data from the Fermi satellite are added.
Deep follow-up observations are triggered at exceptional activity states or at particular epochs
including
XMM-Newton, NuSTAR, and spectroscopy at ground-based optical telescopes. 
A few single observations are conducted  quasi-simultaneous with the Event Horizon Telescope \citep[EHT;][]{EHT2019} observations of OJ 287. 

MOMO represents the densest long-term monitoring of OJ 287 involving X-rays and broad-band SEDs to date. The monitoring cadence is as short as 1 day in cases of outbursts or other noteworthy flux, spectral, or polarimetry states of OJ 287, and is typically 3--4 days in the X-ray--UV--optical bands, and three weeks in the radio bands. 
The theoretical part of the project aims at understanding jet and accretion physics of the blazar central engine in general and the supermassive binary black hole scenario in particular. 

Results are presented in a sequence of publications and so far included: (1) the detection and detailed analysis of two major non-thermal X-ray--UV--optical outbursts in 2016/17 and 2020 \citep{Komossa2017, Komossa2020} and the analysis of the complete long-term Swift light curve at all activity states of OJ 287 \citep{Komossa2021d}; 
(2) Swift, XMM-Newton and NuSTAR spectroscopy of the 2020 outburst around maximum, clearly establishing the spectral components up to $\sim$ 70 keV, including a giant soft X-ray excess of synchrotron origin and an unexpectedly steep spectrum in the NuSTAR band out to 70 keV \citep{Komossa2020};
(3) interpretation of selected events in the context of the binary black hole scenario of OJ 287 \citep{Komossa2020, Komossa2021b}; 
(4) the detection of highly variable radio polarization during the first year of our Effelsberg monitoring in late 2015 and 2016 \citep{Komossa2015, Myserlis2018};
(5) the identification of characteristic and correlated variability across SEDs \citep{Komossa2017, Komossa2021c};  
(6) XMM-Newton and Swift spectroscopy during EHT campaigns in 2018, and a comprehensive analysis of all XMM-Newton spectra of the last two decades \citep{Komossa2021a}; 
(7) the identification of characteristic optical--UV-X-ray time lags in the range 0--17 days based on discrete cross-correlation functions, and estimates of BLR and torus size of OJ 287 to constrain external Comptonization models \citep{Komossa2021d}. 

All Swift data obtained by us are analyzed within days.
The community is alerted in form of  {\sl{Astronomer's Telegrams}} about noteworthy events like outbursts or deep minimum states of OJ 287 we detect with Swift or with the Effelsberg telescope
(ATel \#8411, \#9629, \#9632, \#10043, \#12086, \#13658, \#13702, \#13785, \#14052, \#15145). 
That way, additional multiwavelength observations can be triggered by the community that are not covered by the MOMO program.

Further details of the project data bases and observation strategies with Swift and with the Effelsberg telescope are given below in the respective subsections. 

\subsection{Project data base: MOMO-radio}

In the radio regime, we are using the 100m Effelsberg radio telescope. It offers multiple advantages: a broad frequency range, a large number of receivers at the secondary focus and a high sensitivity.  
Our monitoring of OJ 287 in the course of the MOMO-radio project started in December 2015. Flux and spectral measurements at a dense cadence of typically three weeks are obtained (program identifications 99-15, 19-16, 12-17, 13-18, 75-19, and 65-20, 70-21), covering frequencies between 2.6 and 44 GHz.    

In the radio regime, a coverage of OJ 287 is possible even at that particular epoch each year when OJ 287 is unobservable with ground-based optical telescopes and with Swift due to its solar proximity. The Effelsberg telescope can still observe sources at projected distances of only a few degrees from the Sun (albeit with some occasional degradation of signal quality). 

Radio emission traces the synchrotron emission components. The ongoing measurements are used to (1) obtain flux densities and polarization and their evolution, including the  time intervals around predicted binary impact flares and after-flares; (2) time the radio high-state(s), especially with respect to the multiwavelength data obtained in the MOMO program; (3) measure the evolution of jet emission and magnetic fields;  
and 
(4) distinguish between different SMBBH scenarios and test new predictions of the best-developed binary model, based on distinctly different predictions in the radio regime for the first and second optical peak of the double peaks and for the after-flares. 
For instance, major optical flares 
will not be accompanied by radio flares if they are thermal in nature{\footnote{except for the epochs, when the secondary SMBH may undergo accretion events in conjunction with launching a short-lived jet \citep{Pihajoki2013b, Dey2021}}}. If instead both optical peaks are of synchrotron origin, then two radio flares are expected with a polarization evolution that follows synchrotron theory.  

\subsection{Gamma-ray band}

The MOMO project was designed with the availability of Fermi LAT (Large Area Telescope) $\gamma$-ray data in mind. Unlike the Swift and Effelsberg observations that are proposed and analyzed by us, the $\gamma$-ray data were retrieved from the Fermi archive. They extend the light curve and SED coverage of OJ 287 into the 0.1--100 GeV regime. 
Publicly available Fermi LAT \citep{Atwood2009} data of OJ 287 in the $\gamma$-ray band were retrieved from the Fermi-LAT light-curve repository \citep{Kocevski2021}{\footnote{\url{https://fermi.gsfc.nasa.gov/ssc/data/access/lat/LightCurveRepository/}}}. Weekly averages of the fluxes, and the spectral model with a fixed photon index of 2.16 of a logarithmic parabolic power-law model (logpar) description, were used. 
This description is preferred in our case over variants of free index fits \citep[e.g.][]{Hodgson2017, Kapanadze2018}, because OJ 287 remains in a low $\gamma$-ray state most of the time during the epoch of interest, and fits with free index come with large errors that can introduce spurious luminosity variations in low-states. For any flare states mentioned in the text, we have checked that their identification remains robust when using a free-index fit instead.

\subsection{MOMO-UO and MOMO-X}

To cover the optical to X-ray bands, we have used the versatile space mission Swift because of its broad-band coverage, its high sensitivity, and its scheduling flexibility and fast response time \citep{Gehrels2004}. 

High-cadence light curves and SEDs of OJ 287 are obtained at a cadence that is denser during outbursts (1-3 days), and sparser during more quiescent epochs (3-7 days) with longer gaps when OJ 287 remained constant for several subsequent observations. 
Occasional gaps in the cadence can also be due to the scheduling of a higher-priority target (mostly GRBs), and gaps also arise when OJ 287 is unobservable with Swift due to its close proximity to the moon ($\sim$3--4 days each) or the Sun ($\sim$3 months each year). 
Exposure times are in the range 0.3-2 ks in X-rays - typically 2 ks when OJ 287 was faint, typically 1 ks when it was bright
(Tab. \ref{tab:obs-log}, Fig. \ref{fig:light-CR-Swift}).
Exposure times for the UV--optical telescope (UVOT) are in the same range as the X-ray observations.

While the MOMO project started in late 2015, multiwavelength archival data are added when needed to analyze long-term trends.  
This includes Swift data taken between 2005--2015 \citep{Massaro2008, StrohFalcone2013, Williamson2014, Siejkowski2017, Valtonen2016, Komossa2017} and it includes occasional later Swift data sets not from the MOMO program. 

Here, we present our most recent Swift data of OJ 287 from 2021 and up until 2022 January 15. After that date, Swift went into safe mode for a month. We also add the period 2019--2020 \citep{Komossa2020, Komossa2021d}, for joint analysis with the Effelsberg radio data and Fermi $\gamma$-ray light curve. 
Further details of the data acquisition and analysis in each waveband are reported in the next Sections. 

\subsection{Additional Submillimeter Array data} 

1.3 mm (225 GHz) flux density data were obtained at the Submillimeter Array (SMA) near the summit of Maunakea (Hawaii).
These observations are not part of the MOMO program, but are added here to extend the radio observations to higher frequencies. 
OJ 287 is included in an ongoing monitoring program at the SMA to determine the ﬂuxes of compact extragalactic radio sources that can be used as calibrators at mm wavelengths \citep{Gurwell2007}.  Observations of available potential calibrators are from time to time observed for 3 to 5 minutes, with the measured source signal strength calibrated against known standards, typically solar system objects (Titan, Uranus, Neptune, or Callisto).  Data from this program are updated regularly and are available at the SMA website{\footnote{http://sma1.sma.hawaii.edu/callist/callist.html}}.

\section{Swift data analysis and results}

\begin{table}
\footnotesize
	\centering
	\caption{Log of our Swift observations between 2019 January and 2022 January 15 with observation ids (OBSIDs) 34934-174 to 35905-187. The central wavelengths of the UVOT filters \citep{Poole2008} are reported in the third column and the durations of the single-epoch observations are given in the fourth column. 
	}
	\label{tab:obs-log}
	\begin{tabular}{lclc}
		\hline
		instrument & filter & waveband/central wavelength & $\Delta t$ (ks)\\
		\hline
		XRT &  & 0.3--10 keV & 0.3--2 \\
        UVOT & W2 & 1928\AA  & 0.12--0.6 \\
             & M2 & 2246\AA & 0.09--0.5 \\
             & W1 & 2600\AA  & 0.06--0.3 \\
             & U & 3465\AA & 0.03--0.16 \\
             & B & 4392\AA & 0.03--0.16 \\
             & V & 5468\AA & 0.03--0.16 \\
		\hline
	\end{tabular}
\end{table}

\begin{figure}
\includegraphics[clip, trim=1.0cm 2.3cm 2.0cm 0.3cm, angle=-90, width=\columnwidth]{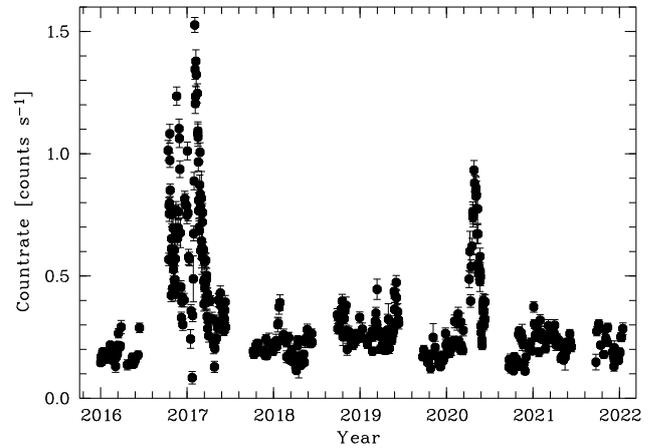}
    \caption{Swift XRT light curve of OJ 287  between 2016 January and 2022 January. 
    }
    \label{fig:light-CR-Swift}
\end{figure}

\begin{figure*}
\includegraphics[clip,width=14cm,trim=1.8cm 5.6cm 1.3cm 2.6cm]{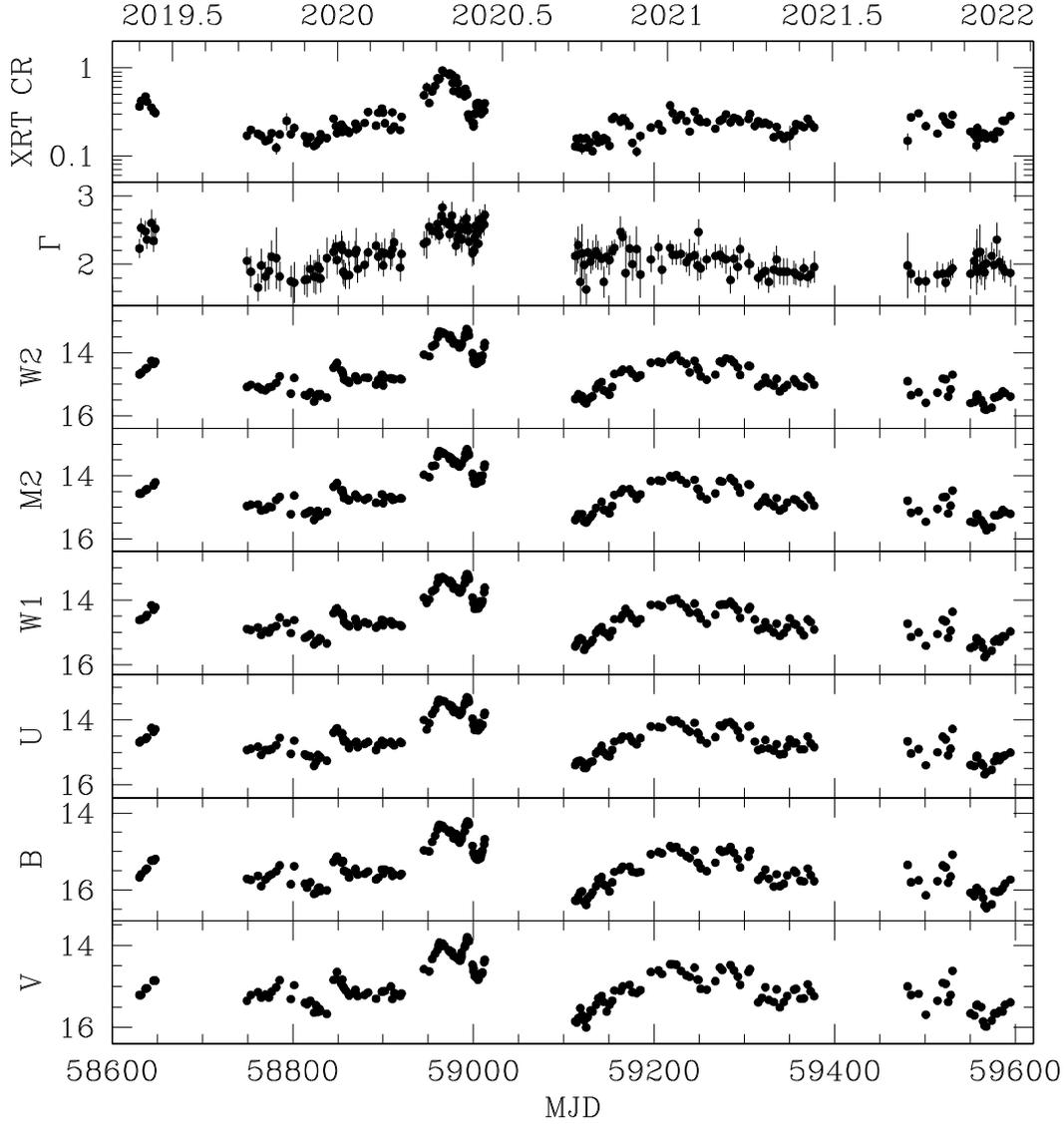}
    \caption{Observed Swift X-ray to optical light curve of OJ 287 since mid-2019 obtained in the course of the MOMO program. The X-ray count rate is reported in cts s$^{-1}$.  The UV--optical filter magnitudes are reported in the VEGA system, and are the directly observed values not yet corrected for Galactic extinction. $\Gamma_{\rm X}$ is the X-ray photon index from single power-law spectral fits. The two long gaps in the light curves correspond to epochs when OJ 287 is unobservable with Swift due to its projected proximity to the Sun.  The last data point is from 2022 January 15. 
   }
\label{fig:lc-Swift-2019-2021}
\end{figure*}

\subsection{Swift XRT data analysis} 

Swift data reduction follows the same procedures we used before \citep[e.g.,][] {Komossa2020, Komossa2021d}. In brief, the X-ray telescope (XRT) data analysis was performed using the {\sc{xrtdas}} package developed at the ASI Science Data Center (SSDC) and included in the {\sc{heasoft}} package (version No. 6.28).
During the majority of the observations, the Swift XRT \citep{Burrows2005} was operating in {\sl{photon counting}} (PC) mode
\citep{Hill2004}. 
Above a source count rate of  
$\sim$1 ct s$^{-1}$, observations were done in {\sl{windowed timing}} (WT) mode \citep{Hill2004}, where only the central 4$\times$4 arcminutes of the CCD are read out. This procedure serves to minimize the effect of photon pileup. 
X-ray count rates in the energy band 0.3--10 keV were determined making use of the XRT product tool at the Swift data center in Leicester \citep{Evans2007}.

To carry out the timing and spectral analysis, we selected source photons within a circular area of a radius of 20 detector pixels. One pixel is equivalent to a scale of 2.36$\arcsec$. 
Because of the required source extraction size, the X-ray jet of OJ 287 detected with Chandra \citep{Marscher2011} is included in the area. 
However, 
at a corresponding Swift XRT count rate  of only 0.009 cts/s,  its contribution to the integrated emission is negligible in all observed X-ray activity states of OJ 287. 
The background photons were extracted in a nearby circular region with a radius of  236\arcsec. 

We note that OJ 287 was off axis in most of the Swift XRT data sets. This is typical for Swift monitoring observations. However, the Swift point-spread function (PSF) does not strongly depend on the location within the inner 
field of view \citep{Moretti2005}. 

The effect of photon pile-up affects data  above a count rate  of $\sim$0.7 cts/s.
To correct for it, we first created a region file where the inner circular 
area of the PSF 
was excluded from the analysis. 
Then, the  loss in counts is corrected by creating a new ancillary response file 
that is used in XSPEC to correct the flux measurement.

X-ray spectra of the source and background in the energy band 0.3-10 keV were generated 
and the spectral analysis was carried out with the package XSPEC \citep[version 12.10.1f;][]{Arnaud1996}. Except when noted otherwise, spectral fits were done with the unbinned data and using 
the W-statistics of \textsc{xspec}. The X-ray count rate light curve of OJ 287 until 2022 January 15 is shown in Fig. \ref{fig:light-CR-Swift}. 

\subsection{Swift UVOT data analysis}

We have also employed the Swift UVOT \citep[UVOT;][]{Roming2005} to observe OJ 287 in all three optical and all three UV photometric bands (see Tab. \ref{tab:obs-log} for the filter central wavelengths).
Here, we focus on data between 2019--2022.04 we have obtained in the course of the MOMO project. The use of all six filters ensures a dense coverage of the SED. This is important since OJ 287 varies rapidly, and epochs of correlated and uncorrelated variability across the optical-to-X-ray bands have been identified previously \citep{Komossa2020, Komossa2021d}.  

Exposure times of the UVOT are in the range 0.3-2 ks. Most of the time, the UVOT bands V:B:U:W1:M2:W2 are observed with a ratio of 1:1:1:2:3:4 of the total exposure time, respectively \citep[e.g.][]{Grupe2010}. 
There are occasional exceptions
when the observation is interrupted by a high-priority target-of-opportunity (ToO) observation. 

For further analysis, the observations in each filter were first co-added, making use of the
tool \textsc{uvotimsum}. In all six filters source counts were then extracted in a region of circular size and with an extraction radius of 5\arcsec centered on OJ 287.  A nearby area of 20\arcsec~radius was used to extract the  
background region.
The tool \textsc{uvotsource} was used to measure magnitudes. 
Background-corrected counts were converted into VEGA magnitudes and 
fluxes making use of the latest calibration \citep{Poole2008, Breeveld2010}.  
All fluxes are reported as flux density multiplied by the central frequency of the corresponding UVOT filter.
For data since 2017, {\sc{caldb}} update version 20200925 was employed.{\footnote{\url{https://www.swift.ac.uk/analysis/uvot/index.php}}}

UVOT data were corrected for Galactic reddening
based on the reddening curves of \citet{Cardelli1989} and using  $E_{\rm{(B-V)}}$=0.0248 \citep{Schlegel1998} and a correction factor in each filter according to Eq. (2) of \citet{Roming2009}. 

\subsection{X-ray spectra}

Spectra were fit with single power laws of photon index $\Gamma_{\rm X}$ (defined as $N(E) \propto E^{- \Gamma}$), 
adding Galactic absorption with a column density $N_{\rm H, Gal}=2.49\times10^{20}$ cm$^{-2}$ \citep{Kalberla2005} and using the absorption model \textsc{tbabs} \citep{Wilms2000}. 
A single power law is the optimal spectral model to analyze single-epoch Swift data as the short observations do not well constrain multi-component models, and the power-law description has been well established as a reliable measure of the X-ray spectral state of OJ 287 below 10 keV.
Further, detailed spectral analysis of all high-sensitivity XMM-Newton data of OJ 287 \citep{Komossa2021a} has shown that excess cold absorption is not required to fit the X-ray spectra of OJ 287.  

Between 2019 and 2021, photon indices are in the range $\Gamma_{\rm X}$=1.7 to 2.8 (Fig. \ref{fig:lc-Swift-2019-2021}) with a trend of steeper indices as OJ 287 becomes X-ray brighter during high-states; a trend that was recognized before \citep[e.g.,][]{Komossa2017, Komossa2021d}. A possible exception is the epoch in 2021 December where $\Gamma_{\rm X}$ shows an indication of steepening while the count rate decreases. However, this trend needs to be confirmed in ongoing monitoring, as measurement errors of single-epoch photon indices are large. 

\subsection{Swift light curves} 

The UVOT light curve during late 2020 to 2021 (Fig. \ref{fig:lc-Swift-2019-2021}) shows a long-lasting systematic rise and fade (referred to as "broad flare"), starting in 2020 September from a low-state and with a first maximum in 2021 January and a second in March, then declining on the timescale of months. Mini-flaring of small amplitude is superposed. While the amplitude of the event does not reach that of the 2020 April--June outburst, the event is of much longer duration.
After the Swift Sun constraint, from 2021 September to November, high-amplitude mini-flaring ($>$ 1 mag in the UV--optical) is then followed by a deep low-state in the UV--optical in December with fluxes as low as 16 mag in the UV (W2) and the optical (V); $\sim$3 mag fainter than during the peak of the 2020 April-June outburst.  

\subsection{Discrete correlation function (DCF)} 

Since the period 2020 September to 2021 June shows an interesting broad flare with mini-flare sub-structure in the UV--optical, we have expanded our previous DCF analysis that covered the  years 2015-2020 \citep{Komossa2021d} and now apply it in the same way to the latest epoch. The time interval 2020 September to 2021 June, between two Swift Sun constraints, is analyzed and we search for correlations between the optical V band and the UV W2 band.

The DCF technique is used to analyze unevenly sampled data sets 
\citep{Edelson1988}. 
We calculated the DCFs based on the  prescription of \citet{Edelson1988} using the 
\textsc{R} package \texttt{sour}\footnote{Available at https://github.com/svdataman/sour.}
\citep{Edelson2017}. 
The time step $\tau$ over which the DCFs were computed corresponds 
to twice the median time step across the September--June light curve.
To evaluate the significance level of the measured lags, confidence contours for each DCF were produced by simulating 
$N=10^3$ artificial W2 light curves. The prescription of \citet{Timmer1995} was adopted, assuming a power spectral density 
of $P\left(f\right) \propto f^{-\alpha} = f^{-3}$ based on the results of 
the structure function (SF) analysis for W2 \citep{Komossa2021d}, taking $\alpha = \beta + 1$, where $\beta$ is the SF slope. 
Artificial DCFs were then computed based on these artificial light curves, 
allowing for the computation of the 
$90^{\mathrm{th}}$, $95^{\mathrm{th}}$, and $99^{\mathrm{th}}$ percentiles based on the distribution of artificial DCFs at each time step. 

\begin{figure}
\includegraphics[width=\columnwidth]{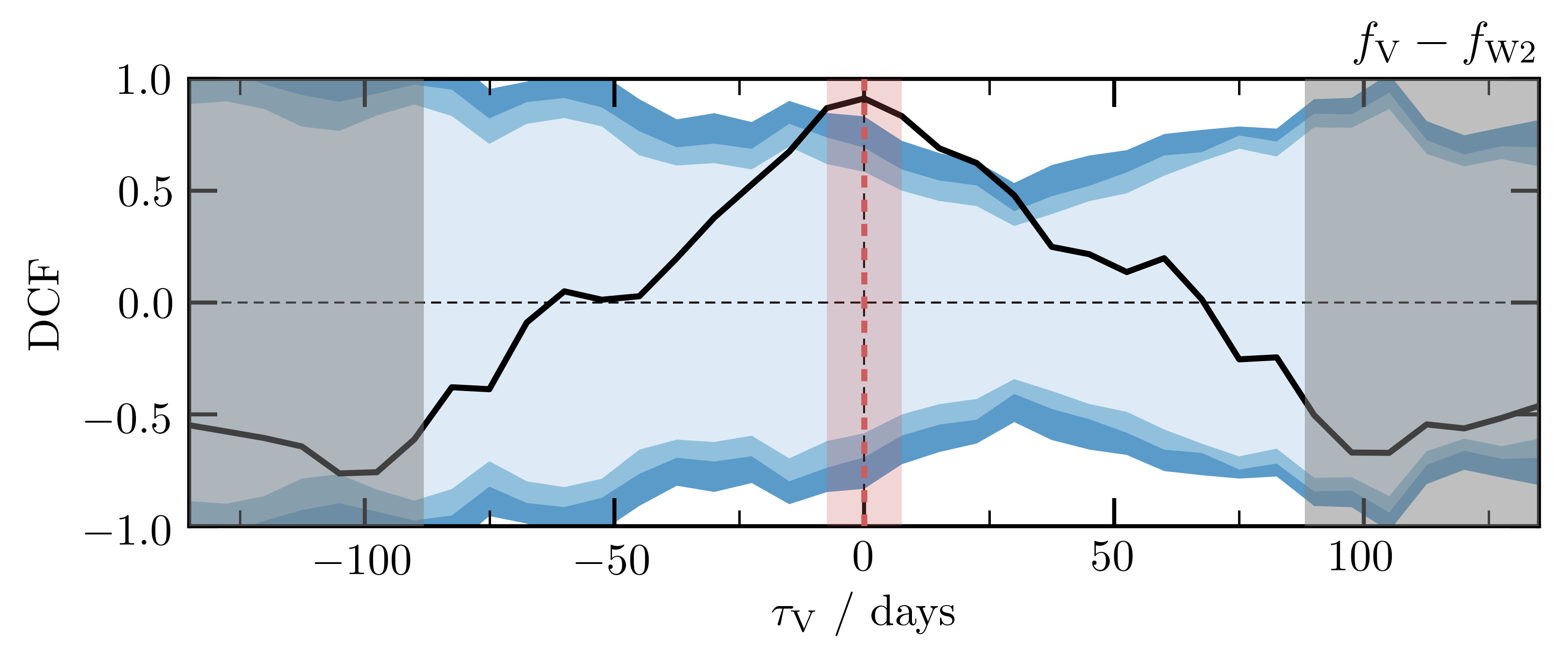}
\caption{V$-$W2 DCF of OJ 287 for the epoch 2020 September to 2021 June (black line).  
Filled regions indicate the $\pm90^{\mathrm{th}}$ 
(light blue), $\pm95^{\mathrm{th}}$ (blue), and $\pm99^{\mathrm{th}}$ (dark blue) percentiles from the 
light-curve simulations.
The horizontal dashed line 
marks zero correlation 
and the vertical dashed line indicates $\tau_{\mathrm{V}} = 0~\mathrm{days}$. Positive $\tau_{\mathrm{V}}$ values indicate V leading W2, negative values indicate lagging.
Gray regions are between one-third to one-half the total light-curve length where results become more unreliable. 
The vertical red line marks the measured time lag and its error. 
UV and optical fluxes are closely correlated with a lag consistent with zero days.   
 } 
\label{fig:DCF}
\end{figure}

To evaluate the error of the lag measurements, the autocorrelation function (ACF) of the W2 light curve was computed, 
following the same procedure described above to create confidence contours for the ACF. 
Because the ACF peaks at 
$\tau = 0~\mathrm{days}$ we estimated the error on the lag measurement as all ACF values in excess of 
the $99^{\mathrm{th}}$ percentile contour around $\tau = 0~\mathrm{days}$. 
Measured lags are determined as those times where the DCF exceeds the $99^{\mathrm{th}}$ percentile contour for 
either an anti-correlation or a correlation. The times for which a lag measurement is reported are restricted to times corresponding to $\leq 1/3$ the length of each light-curve. 
We find that the V and W2 light curves are closely correlated with a lag of 0 $\pm$ 7.5 days (Fig. \ref{fig:DCF}). 

\subsection{Fractional rms variability amplitude}

\begin{table}
	\centering
	\caption{Fractional variability amplitude $F_{\rm var}$ of the Swift UVOT and XRT fluxes of OJ 287 during 2020--2021.}
	\label{tab:Fvar}
	\begin{tabular}{lcc} 
		\hline
		 & Epoch 6 & Epoch 7 \\
		 & 2020 September -- 2021 June & 2021 September--December \\
		\hline
$f_{\mathrm{x}}$ & $0.254\pm0.011$ & $0.28\pm0.02$ \\ 
$f_{\mathrm{W2}}$ & $0.375\pm0.005$ & $0.32\pm0.01$ \\ 
$f_{\mathrm{M2}}$ & $0.380\pm0.006$ & $0.34\pm0.01$ \\ 
$f_{\mathrm{W1}}$ & $0.374\pm0.005$ & $0.36\pm0.01$ \\ 
$f_{\mathrm{U}}$ & $0.369\pm0.005$ & $0.38\pm0.01$ \\ 
$f_{\mathrm{B}}$ & $0.374\pm0.004$ & $0.37\pm0.01$ \\  
$f_{\mathrm{V}}$ & $0.362\pm0.006$ & $0.37\pm0.01$ \\ 
		\hline
	\end{tabular}
\end{table}

The fractional rms variability amplitude $F_{\rm var}$ \citep{Vaughan2003} was computed for the two most recent Swift epochs, separately for each
UVOT filter and in the X-ray band. 
$F_{\rm var}$ is defined as
\begin{equation}
F_{\rm var} = \sqrt{S^2 - \overline{\sigma^{2}_{\rm err}} \over \overline{x}^2},
\end{equation}
where $S^{2}$ is the variance of the light curve, $\overline{\sigma^{2}_{\rm err}}$ is the mean square of the
measurement uncertainties, and $\overline{x}$ is the mean flux. The error of $F_{\rm var}$ was computed following Appendix B of \citet{Vaughan2003} as
\begin{equation}
\sigma_{F_{\rm var}} = \sqrt{ \left(\sqrt{1 \over 2N} {\sigma^{2}_{\rm err} \over \overline{x}^2 F_{\rm var}} \right)^2 + \left( \sqrt{\sigma^{2}_{\rm err} \over N} {1 \over \overline{x}^2} \right)^2 },
\end{equation}
where $N$ is the number of data points used in the computation of $F_{\rm var}$. 

The fractional variability is similar in all optical and UV bands and slightly enhanced w.r.t. the X-ray band.
Results are reported in Tab. \ref{tab:Fvar}. 

\section{Effelsberg multi-frequency radio observations}

\subsection{Data acquisition and reduction}

\begin{table*}
	\centering
	\caption{Receivers used in our Effelsberg observations of OJ 287 since 2019 (Effelsberg program IDs 13-18, 75-19, 65-20, and 70-21). 
    Only the radio frequencies most recently employed since 2019 are listed. (Some receivers have changed in the past,
     and at selected epochs a larger number of frequencies was observed in the course of the MOMO program.) $\nu_{\rm center}$ is the central frequency, $\Delta\nu$ the band width and HPBW the half power beam width.  
	}
	\label{tab:obs-log-radio}
\begin{tabular}{ccccl}
\hline
Receiver & $\nu_{\rm center}$  & $\Delta \nu$ & HPBW & Comment\\
 &  [GHz] &  [MHz] & [arcsec] & \\
\hline
S110mm & 2.595 & 10 & 286 \\
S60mm & 4.85 & 500 & 150 \\
S28mm & 10.45 & 300 & 67.5 \\
S20mm & 14.25 & 2500 & 52.9 \\
    & 16.75 & 2500 & 43.7\\
S14mm & 19.25 & 2500 & 40.1 & S14 included from 2020 April\\
    & 21.15 & 2500 & 38.0 & occasional RFI at 19.25 GHz\\
    & 22.85 & 2500 & 36.8 & \\
    & 24.75 & 2500 & 33.1 & \\
S7mm  & 36.25 & 2500 & 23.0 & 35.75 GHz until MJD 58981 (2020 May 12) \\
    & 38.75 & 2500 & 21.2 & 38.25 GHz until MJD 58981 (2020 May 12) \\
    & 41.25 & 2500 & 20.7 & used since 2021 March 24\\
    & 43.75 & 2500 & 19.7 & used since 2021 March 24\\
\hline
\end{tabular}
\end{table*}

\begin{figure*}
\includegraphics[clip, trim=0.9cm 5.3cm 1.6cm 8.4cm, width=13cm]{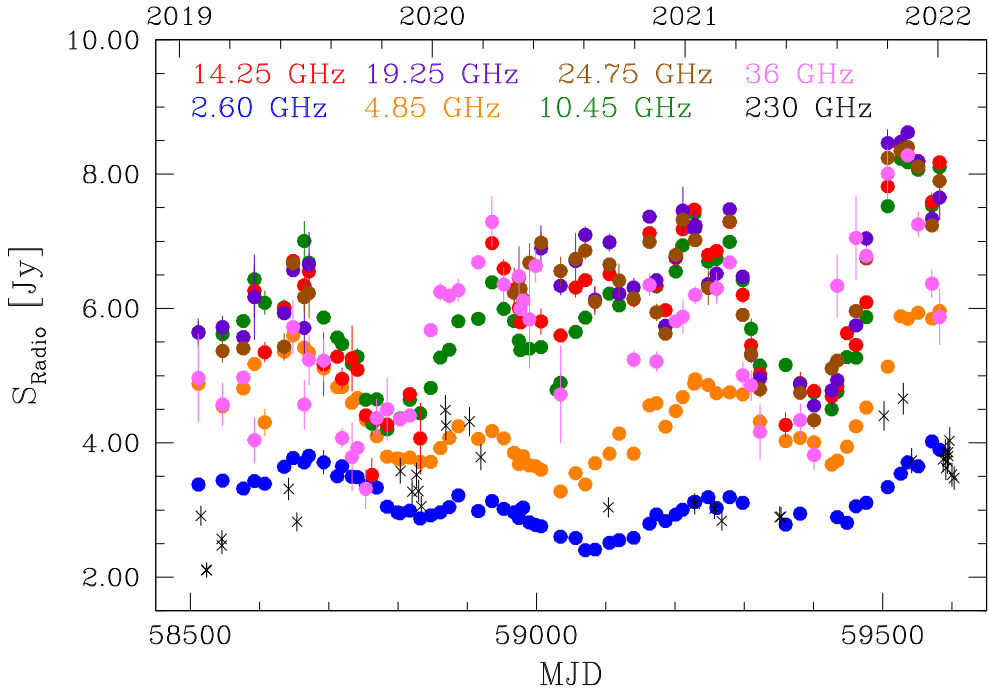}
	\caption{Radio light curves of OJ 287 between 2019 and 2021 measured at the Effelsberg telescope and with the SMA.
     }
     \label{fig:EB_radio_lc}
\end{figure*}

Observations were obtained  between 2.6 and 44 GHz  switching between up to six 
receivers (Tab. \ref{tab:obs-log-radio}). The cadence of observations was 3--4 weeks, or higher at selected epochs. Depending on weather conditions, the coverage at the highest frequencies was more sparse.

We note that the coverage of OJ 287 in mid-2019 (late July to August) was affected by the close proximity of OJ 287 to the Sun.
In particular, at the time closest to the predicted binary impact flare and the Spitzer monitoring \citep{Laine2020}, the Effelsberg observation of OJ 287 could not be conducted because the painting of the structural parts was renewed. With the paint just removed and OJ 287 only a few degrees away from the Sun at the time interval in question, the extra reflectivity of the telescope support structure hindered the observation. 

The cross-scan method \citep{Heeschen1987, Kraus2003} was used to acquire the radio data.
In the cross-scans, the telescope was moved in two perpendicular directions, azimuth and elevation, with the target source position at the center of the scans. 
The number of sub-scans varied between 2-3 per direction for the low and 12-16 for the high frequencies.
The observing time needed to measure the flux densities at all frequencies once was 25 minutes. 

The data reduction and analysis was performed in a standard manner as described in, e.g., \citet{Kraus2003}. 
In a first analysis step, 
a Gaussian profile was fit to the data of every single sub-scan. 
Bad sub-scans (for instance due to high pointing errors, radio frequency interference (RFI), or --- in case of OJ 287 --- disturbances by solar radiation around August 1st of each year) were identified and excluded from further analysis. 
After correcting for small pointing errors of the telescope, the amplitudes of the individual sub-scans were averaged. In some cases, at the highest frequencies and in mediocre
weather conditions, the sub-scans of one direction (azimuth/elevation) were averaged before the fitting of the Gaussian profile to increase the signal to noise ratio.
Next, corrections for the atmosphere’s opacity were applied as well as for the gain-elevation effect (change of sensitivity with elevation). 
Finally, absolute flux calibration was achieved by comparing the observed antenna temperatures with the expected flux densities of selected calibrators like 3C 286  {\footnote{We note in passing that the radio calibrator 3C 286 has recently shown variability in the $\gamma$-ray and X-ray band \citep{YaoKomossa2021} opening the possibility of radio-variability of the inner jet, too. However, the bulk of the radio emission of 3C 286 is widely extended and 3C 286 is at high redshift, and therefore low-resolution radio observations as the ones carried out here will be unaffected by any variability of the inner jet emission.}}. A detailed description of the analysis procedures will be given by Kraus et al. (in preparation).

\begin{table*}
	\centering
	\caption{Summary of the properties of the four radio flares at peak. Column entries: (1) Flare designation, (2) peak flux density at 36 GHz, (3) modified Julian date (MJD) of the 36 GHz peak (uncertain by $\pm$ 10 d given the observing cadence), (4) calendar date of the 36 GHz peak, (5) MJD of the optical peak, (6) time difference between radio (36 GHz) and optical ($V$) peak in days, (7) ratio of peak fluxes in the optical ($V$) and radio (36 GHz) band, (8) spectral index $\alpha_{\nu}$ at peak (between 10 and 36 GHz), (9) spectral index $\alpha_{\nu}$ between 36 GHz and 225 GHz (observations at these two frequencies are not simultaneous, but agree within 1-7 days except for 16 days at F2). 
	}
	\label{tab:flare-summary}
	\begin{tabular}{lllllclcc} 
		\hline
Flare &  $S_{\rm peak, 36 GHz}$ & MJD$_{\rm 36GHz_{peak}}$ &  date$_{\rm 36\,GHz_{peak}}$ & MJD$_{\rm V_{peak}}$ & $\Delta t_{\rm{V-36\,GHz}}$ & $f_{\rm V}$/$f_{\rm 36\,GHz}$ & $\alpha_{\nu,\rm{10-36\,GHz}}$ & $\alpha_{\nu,\rm{36-225\,GHz}}$\\
		 & [Jy] & &  &  & [d] & & & \\
            (1)  & (2) & (3) & (4) & (5) & (6) & (7) & (8) & (9) \\
		\hline
F1 & 5.72$\pm$0.40 & 58648.557 & 2019--06--14  & 58647.964 & $-0.6$ & 12.2$^1$ & $-$0.12 & $-$0.30\\
F2 & 7.29$\pm$0.38 & 58935.896 & 2020--03--27 &  58962.351 & +26.5 & 22.5$^2$ & +0.11 & $-$0.36\\
F3 & 6.20$\pm$0.03& 59229.138 & 2021--01--15 & 59220.480 & $-8.7$ & 16.3$^3$ & $-$0.12 & $-$0.38\\
 & 6.69$\pm$0.05& 59278.991 & 2021--03--05 & 59284.463 & +5.5 & 14.8 & $-$0.03 \\
F4 & 8.28$\pm$0.11& 59536.240 & 2021--11-18 & 59530.448$^4$ & $-5.8$& -- & +0.01 & $-$0.31\\  
		\hline
\end{tabular}

{Notes: $^1$The last observed $V$ flux was used for this ratio, since OJ 287 entered Swift Sun constraint afterwards. 
$^2$ MJD and flux ratio correspond to the first optical peak of the triple-peaked optical outburst. 
$^3$ This broad flare has two optical peaks of similar flux, and both peaks are listed.
$^4$At this epoch, no clear flare can be defined in the optical--UV. The flux shows some rapid variations by a factor $>$2. The MJD corresponds to the single highest $V$ flux.  } 
\end{table*}

The measurement uncertainties are based on the errors resulting from the
least-squares-fit of the Gaussian profiles and statistical errors from averaging
of the data. These errors are propagated throughout the data reduction
process and combined with a final contribution which reflects the apparent
residual fluctuations of the calibrators.
At frequencies below 15 GHz, the final relative uncertainties are usually
well below 5\%. The errors increase at higher frequencies due to the increased
influence of weather effects, but are mostly of the order of 5-10\%.

\begin{figure}
\centering
\includegraphics[clip, trim=1.0cm 2.0cm 1.3cm 0.3cm, angle=-90, width=7.5cm]{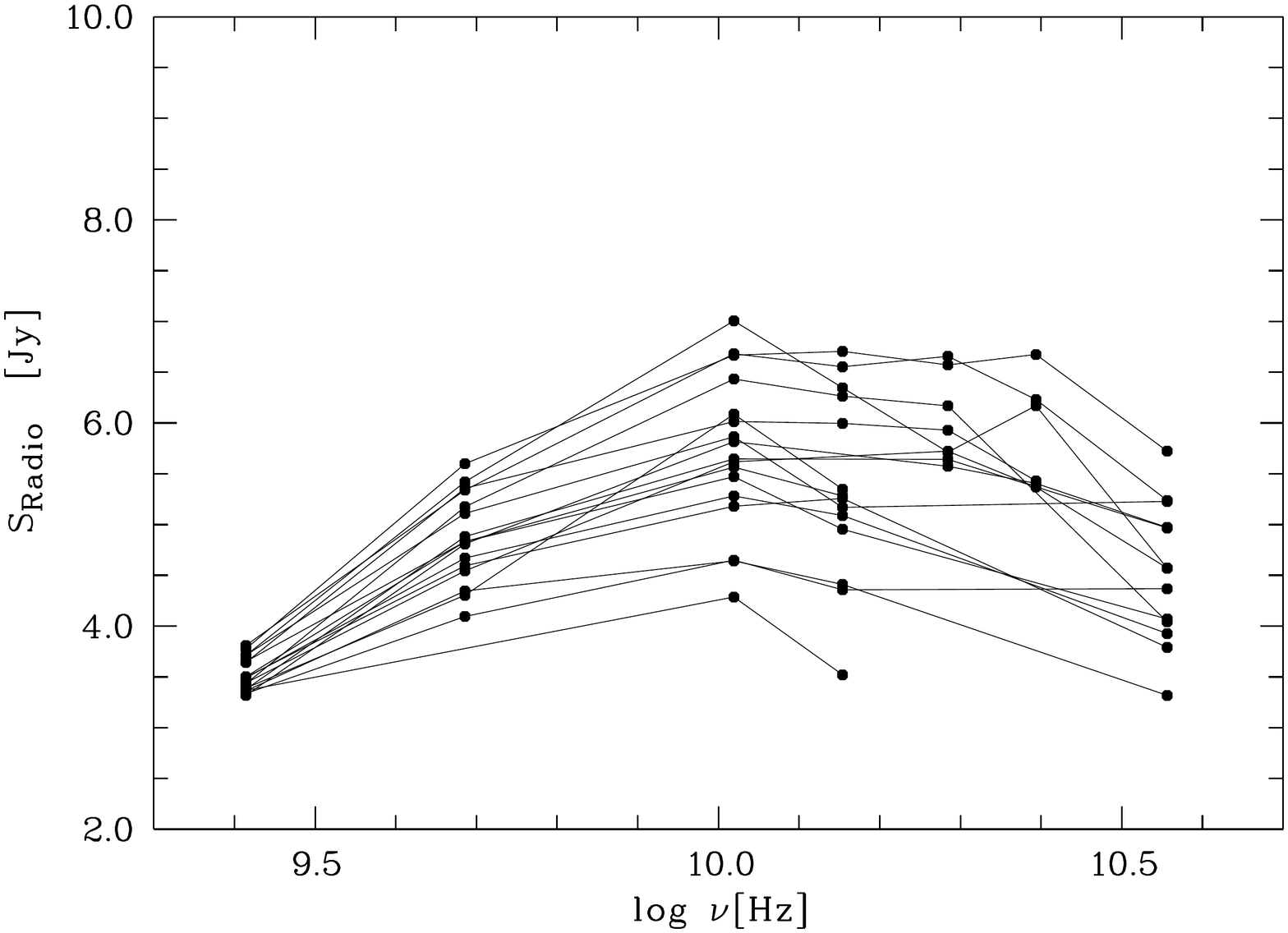}
\includegraphics[clip, trim=1.0cm 2.0cm 1.3cm 0.3cm, angle=-90, width=7.5cm]{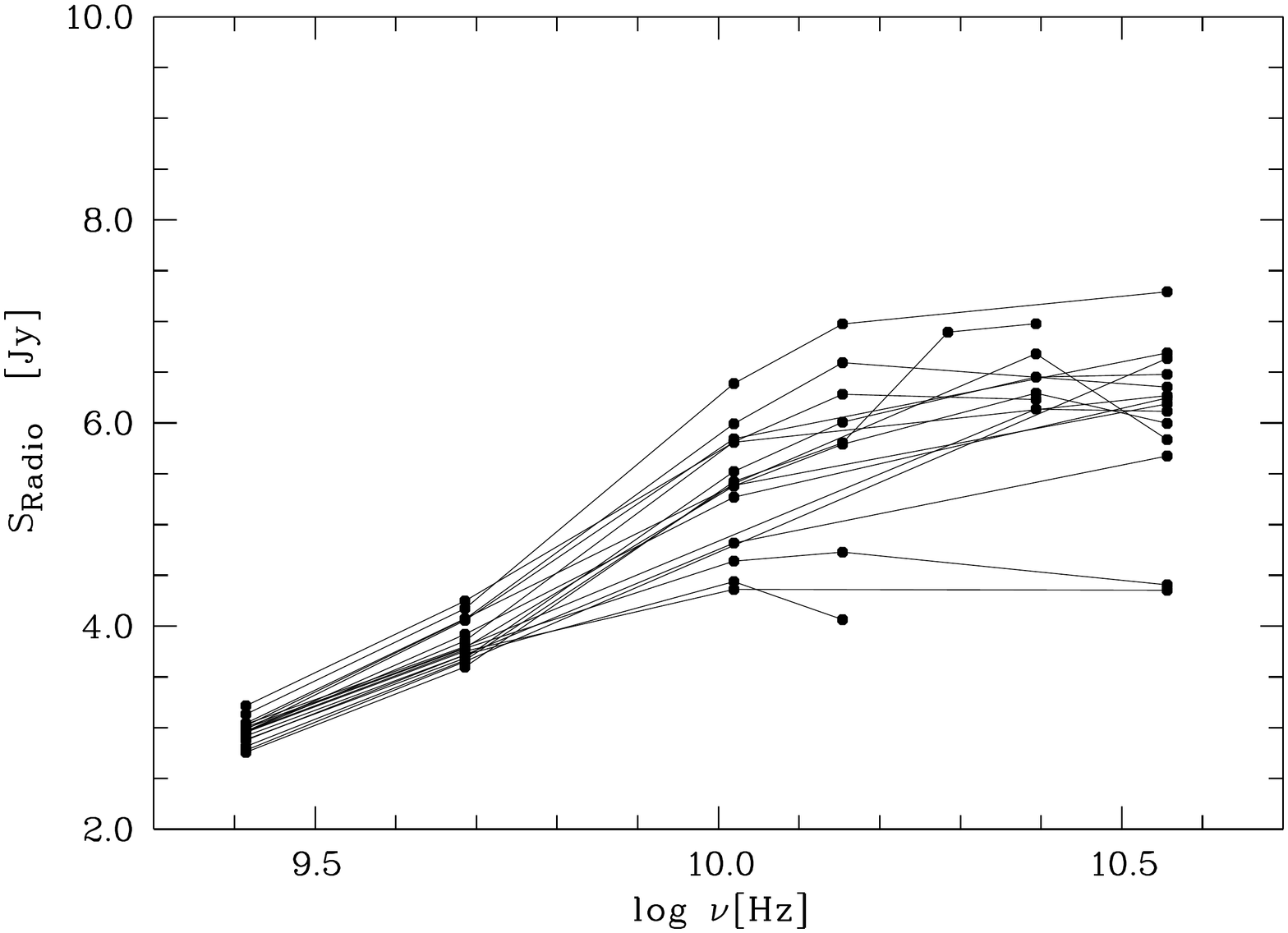}
\includegraphics[clip, trim=1.0cm 2.0cm 1.3cm 0.3cm, angle=-90, width=7.5cm]{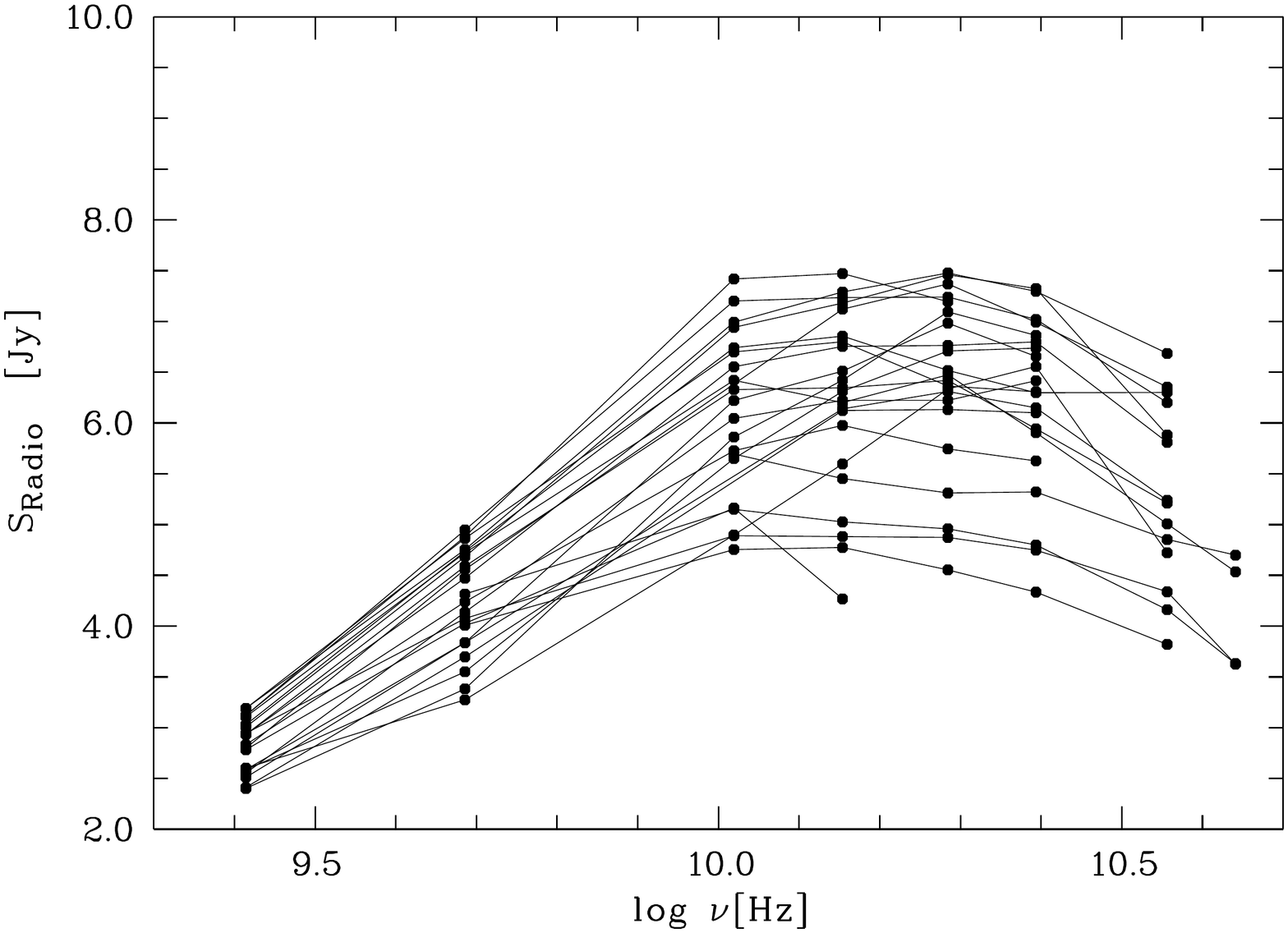}
\includegraphics[clip, trim=1.0cm 2.0cm 1.3cm 0.3cm, angle=-90, width=7.5cm]{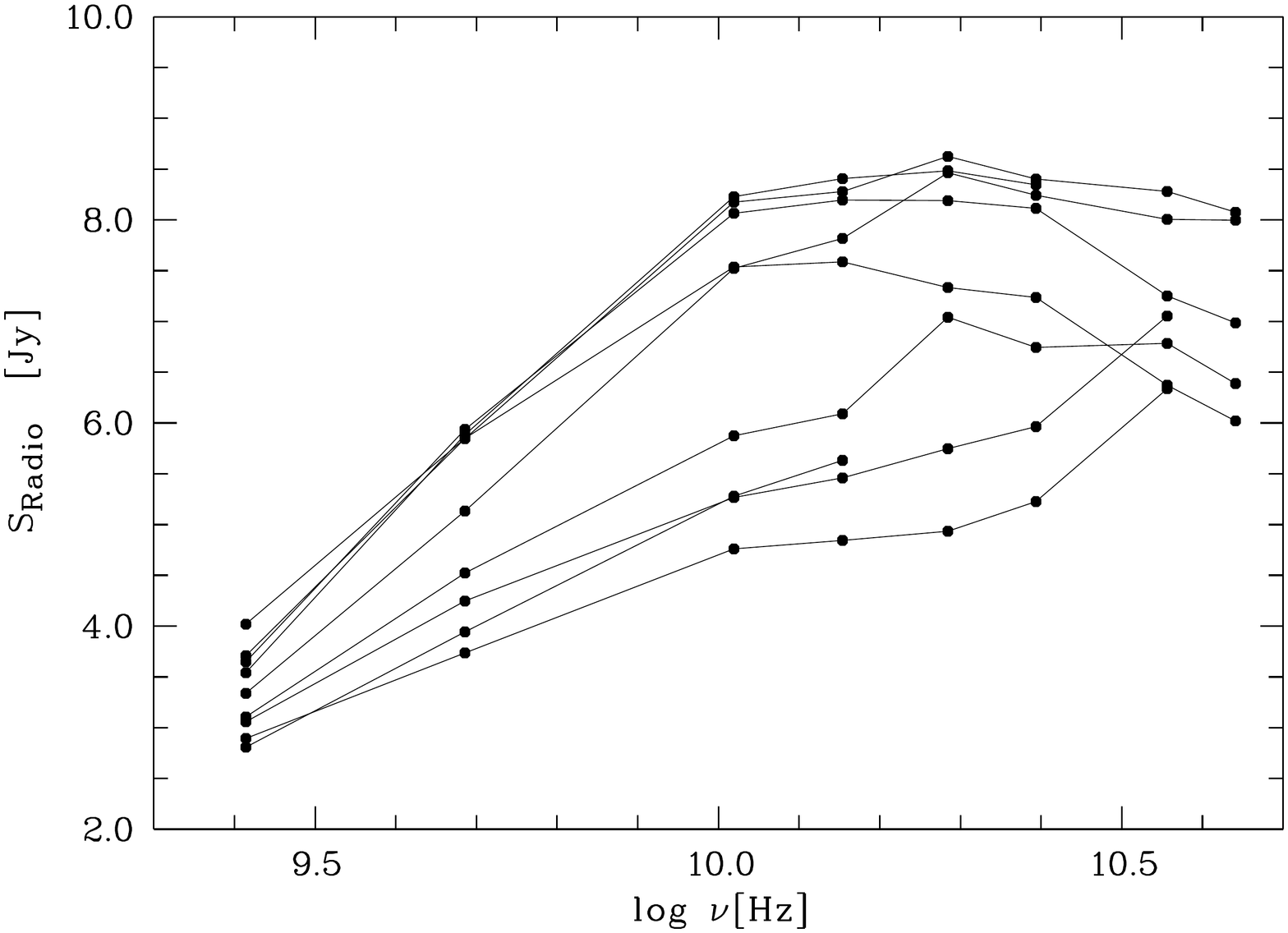}
	\caption{Radio SEDs of OJ 287 between 2019 and 2021  (upper panel 1: MJD 58511 -- 58784, panel 2: MJD 58799 -- 59028, panel 3: MJD 59034 -- 59426, panel 4: MJD 59434 -- 59582). Each panel includes one of the flares. 
     }
     \label{fig:EB_radio_spectra}
\end{figure}

\subsection{Radio light curves} 

OJ 287 is found at a high level of activity in the radio band. 
Radio light curves (Fig. \ref{fig:EB_radio_lc}) show four separate main maxima in 2019 June, 2020 March, 2021 March, and 2021 November (dates are based on observations at $\nu$=36 GHz; Tab.\ref{tab:flare-summary}). 
The maxima are detected at all frequencies but are more pronounced at higher frequencies where sometimes additional substructure is evident.
Based on the highest and lowest flux density measured during the whole epoch of observations, the total amplitude of variability is a factor of 1.67 (2.6 GHz), 1.96 (10.45 GHz), and 2.50 (36 GHz). 

The single sharpest and the brightest flare, starting to rise in late July 2021, approximately doubled its flux density in 135 days between $S$ = 3.82 Jy at low-state and 8.28 Jy at maximum (36 GHz).
The flare maximum was reached at the highest frequencies first. After the peak in 2021 November the flux densities declined in December, but remained at high emission levels. 

\subsection{Radio SEDs}

Radio SEDs are displayed in Fig. \ref{fig:EB_radio_spectra}. They are highly variable and show a range of turn-over frequencies between 10 and 25 GHz. Radio spectral indices, defined as $f_{\nu} \propto \nu^{\alpha_\nu}$, vary between $\alpha_\nu$ = 0.17 to 0.65 (in the band 2.6-10.45 GHz) and between $\alpha_\nu = -0.38$ to 0.23 (10.45-36 GHz). 

In mid-2019 an inversion of the radio spectrum is seen where high frequencies show similar flux density levels as those at 4.85 GHz. Flux densities at 36 GHz reach particular low values. 

\subsection{Broad-band light curves}

\begin{figure*}
\includegraphics[clip, trim=0.6cm 5.4cm 1.3cm 2.6cm, width=12cm]{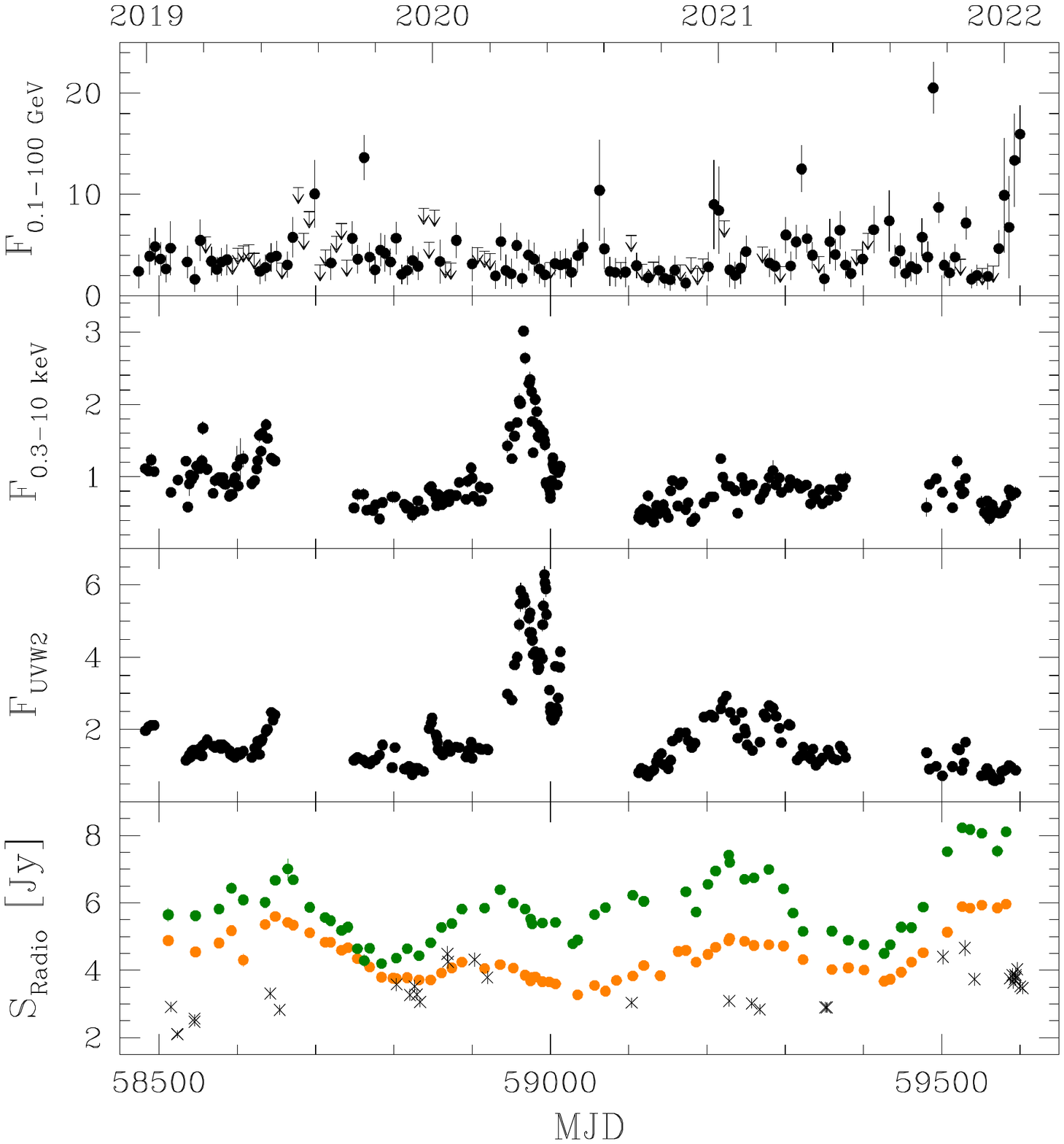}
    \caption{OJ 287 light curves from 2019 to 2022.04 in $\gamma$-rays, X-rays, UV-W2, and selected radio frequencies (orange: 4.85 GHz, green: 10.45 GHz, black: 225 GHz).
    The $\gamma$-ray flux, observed absorption-corrected X-ray flux (0.3-10 keV) and the extinction-corrected optical--UV fluxes are given in units of 10$^{-11}$ erg s$^{-1}$ cm$^{-2}$. The radio flux density is given in Jy.
 }
\label{fig:lc-MWL}
\end{figure*}

Figure \ref{fig:lc-MWL} compares the Effelsberg, SMA, Swift and Fermi light curves between 2019 and 2022 January.  Weekly averages of the Fermi fluxes were used and constant index of the logarithmic parabolic model (see Sect 2.3) was assumed.  
The first radio flare falls within an epoch when OJ 287 was unobservable with Swift due to its solar proximity. However, one can clearly see the rise in flux in the Swift UVOT bands; likely the onset of a flare the peak of which escaped detection. 
The radio light curve shows a second flare that accompanies the 2020
outburst detected with Swift. A third radio flare of comparable peak flux is also seen in 2021, when the X-rays remained in a rather low state and the optical--UV showed mini-flares superposed on a
broader multi-months flare with intermediate peak flux. The two radio flares overlapped to form a broad emission hump. 
A further, well separated, sharp radio flare is seen in late 2021, reaching the highest flux levels during the whole 2019--2021 time interval of observations. While the optical--UV and X-rays showed a deep low-state throughout 2021 December, the radio flux levels remained high.

While the peak radio flux density of the flares is rather similar (values between 6--8 Jy), the peak ratio of the optical over the radio flux varies more strongly among the flares. The second flare has the highest relative optical flux, while the last one has the highest relative radio flux with no particular longer-lasting optical flare event identifiable (Tab. \ref{tab:flare-summary}).

Like the radio observations, the Fermi $\gamma$-ray light curve is continuous as well, in contrast to ground-based optical and space-based Swift data that come with a $\sim$3 month gap due to OJ 287's solar proximity every year. 
Overall, the Fermi-LAT $\gamma$-ray light curve shows a much lower level of activity since 2017 (Fig. \ref{fig:lc-Swift-signature}) than it showed before. The epoch pre-2017 was characterized by several bright and broad $\gamma$-ray flares \citep[e.g.,][]{Abdo2009, Hodgson2017} 
with evidence for correlated variability in the radio (millimeter-wave) and $\gamma$-ray band \citep{Agudo2011}. 
In particular, there is no evidence that the 2020 outburst detected with Swift in all wavebands does have a near-simultaneous bright $\gamma$-ray counterpart.
Since 2019 (Fig. \ref{fig:lc-MWL}), OJ 287 has been relatively inactive at $\gamma$-rays.  
The highest post-2016 $\gamma$-ray flux was reached around
2021 October 3
at a flux level of 2.11 ($\pm$ 0.26) $^{-4}$ MeV cm$^{-2}$ s$^{-1}$ (three-day average).    
Given the findings of \citet{Agudo2011} it is possible that the bright $\gamma$-ray flare is associated with the strong radio flare that peaks in mid-November, but ongoing observations of similar events are needed to evaluate their correlation. 

\section{Discussion} 

\subsection{Activity state of OJ 287 in the radio band}

We have caught OJ 287 in a state of high activity in the radio band during the period 2019-2022.04 with several bright flares. The broad-band SED from radio to X-rays is shown in Fig. \ref{fig:SEDs_broad}. 

Radio spectral indices are in the range $\alpha_\nu=-0.38$ to 0.23 at high frequencies (10.45-36 GHz) and higher at low frequencies. 
Overall, the turn-over frequencies of the radio SED have shifted to lower frequencies (10-25 GHz) during the epoch of observations, in comparison with \citet{Lee2020} who reported values between 30-50 GHz during the earlier epoch 2013-2016, implying that OJ 287 became optically thin at lower frequencies during its recent activity since 2019. 

The pronounced radio flare at the end of 2021 is used to estimate the apparent brightness temperature and Doppler factor during that epoch. 
At 36.25 GHz, and based on a flux doubling timescale of 135 days, we obtain an apparent brightness temperature of $T_{\rm B,app}=6.7 \times 10^{12}$ K.
Taking the inverse Compton limit of $T=10^{12}$ K as conservative limit \citep{Kellermann1969} and assuming a spectral index $\alpha=-1$ of the optically thin part of the radio spectrum then gives a minimum Doppler factor of $\delta = 3.4$ of the emitting component.  

We come back to the radio emission in comparison with the other wavebands when discussing epochs of particular interest below. 

\subsection{Epochs of special interest}

During the period 2019--2022.04 that is the focus of this study, several remarkable flux and spectral states of OJ 287 were detected. Some of these coincide in time with predictions from the SMBBH model, while others are likely driven by jet physics independent of any binary's presence.  We discuss each of these epochs in turn in the next Sections.
When we compare with predictions of the binary model of OJ 287, these are based on \citet{Lehto1996} with latest numbers from \citet{Dey2018}. In brief, the model predicts impact flares when the secondary SMBH crosses the primary's accretion disk, and after-flares when the impact disturbance reaches the inner disk and triggers changes in jet activity. 
The flares are not equidistant in time, but their timing changes in a  predictable manner, based on the model for the elliptical, precessing orbit of the secondary SMBH. 
The main flares associated with disk impacts do not become observable immediately but only after a time delay corresponding to the time interval it takes the expanding, impact-driven bubble to become optically thin \citep{Ivanov1998, Valtonen2022}. Therefore, below, we distinguish between the time of the actual disk impact event, and the later, major "impact flare". 

Our observations cover two main episodes of predicted SMBBH activity: 
The epoch of the 2019 impact flare that we cover in the radio band (not with Swift, since OJ 287 was in Swift Sun constraint), and the epoch of the 2021 December secondary-SMBH disk crossing (the actual predicted impact flare will only be visible later, after the expanding bubble becomes optically thin). We comment on each in turn, in addition to other epochs of interesting flux/spectral states of OJ 287. 

\subsubsection{Epoch of 2019 July--August and binary impact flare} 

During mid-2019, the epoch of the latest predicted impact flare, OJ 287 was neither observable from the ground in the NIR-optical regime, nor from space with Swift, due to its close proximity to the Sun.
The Spitzer Space Telescope with its Earth-
trailing orbit was used instead and the detection of a (thermal) IR flare at the expected time was reported by \citet{Laine2020}.

Our radio monitoring also covered the larger time frame (2019 July--August) and we observed at a higher cadence.
During the epoch, a broad radio flare (lasting months) was detected. The flare maximum was reached around 2019 June 14 
about six weeks before the IR flare detected with Spitzer with peak on 2019 July 31 (note, however, that the Spitzer flare was interpreted as thermal emission, so the two events are then not directly related).  
During this epoch there is an interesting inversion of the radio spectrum with flux densities at the higher frequencies as low as at 4.85 GHz and particularly low flux densities at 36 GHz. This is likely at least partly due to opacity-dependent time delays for each frequency band to reach its maximum, with higher frequencies turning down faster than lower frequencies, along with rapid variability at the highest frequency. The spectral inversion persists until October 2019.

During the Spitzer IR flare, a short $\gamma$-ray flare was recorded with Fermi; the second-brightest Fermi flare during 2019 (Fig. \ref{fig:lc-MWL}). The $\gamma$-ray flare peaked quasi-simultaneous with the IR flare on 31 July 2019, 
suggesting that the two are related.{\footnote{Even though we do not discuss epochs pre--2019 much further here, it is interesting to note that the previous binary impact flare of December 2015 \citep{Valtonen2016} was accompanied by a  $\gamma$-ray flare, too, simultaneous to within a day (Sect. A, Fig. \ref{fig:lc-Swift-signature}). It was the brightest flare recorded since late 2015. 
{\em If} the secondary possessed a jet during its impact (but see Sect. 5.2.4), then its interaction with the dense accretion disk could be a potential source of $\gamma$-ray emission, in analogy to the $\gamma$-ray production mechanism discussed by \citet{Araudo2010}, who consider jet interactions with dense broad-line region clouds.  
Such a scenario could potentially explain the 2019 $\gamma$-ray flare, but not the 2015 $\gamma$-ray flare because 2.39 yrs had already passed since the actual disk impact according to the model \citep{Dey2018, Valtonen2022}, and the secondary SMBH had already crossed the disk when in 2015 the near-side impact-driven bubble became optically thin and the {\em optical} thermal flare became observable. It is possible, however, that the far-side bubble triggered renewed accretion and jet activity of the secondary's jet with new $\gamma$-ray flaring as a consequence, when interacting with the accretion disk. Mechanisms to produce $\gamma$-ray emission in SMBBH context will be further discussed in future work.
}} 

\subsubsection{2020 April--June outburst}

This non-thermal outburst was reported and discussed in detail by \citet{Komossa2020} including the initial Swift discovery along with dedicated deep follow-up observations with XMM-Newton and NuSTAR and their modelling and interpretation, and by \citet{Komossa2021d} in context with the other OJ 287 flux and spectral states of interest since 2005.
[Our initial report about the detection of this outburst (ATel \#13658) also triggered an additional X-ray observation \citep{Prince2021, Singh2021} that still found OJ 287 in the supersoft X-ray state that characterized this outburst.]  
Here, we only point out that the detection of radio flaring activity independently confirms the previous conclusions about the non-thermal nature of the outburst. In particular, this also supports the previous conclusion that the event is consistent with an after-flare predicted by the binary model \citep{Sundelius1997} where new jet activity is launched.  

\subsubsection{The 2020--2021 broad flare} 

During late 2020 and 2021 a remarkable, long-lasting UV--optical flaring event ("broad flare") is evolving at intermediate emission levels, accompanied by radio flaring. 
The fractional variability amplitude 
$F_{\rm var}$ during this epoch (Tab. \ref{tab:Fvar}) is intermediate between previous states of outbursts (2016/2017 and 2020) and states of quiescence of OJ 287
\citep{Siejkowski2017, Komossa2021d}, and significantly higher than the near-zero variability of OJ 287 on sub-day timescales observed in X-rays with XMM-Newton \citep{Gallant2018, Komossa2020, Komossa2021d}.

During this epoch, the power law photon index $\Gamma_{\rm x}$ is no longer as closely correlated with source count rate as it was in previous epochs, with an indication of an emerging anti-correlation at the end of 2021, as opposed to the strong correlation so far seen during outburst epochs \citep[e.g.,][]{Komossa2021a} driven by a soft synchrotron component. 
An anti-correlation can be understood if a hard IC component makes a stronger contribution to the recent X-ray spectra. 
Future observations are needed to confirm this trend. 
In previous epochs, the UV and optical emission of OJ 287 has been closely correlated with time lags consistent with $\tau=0\pm1$ days at all activity states of OJ 287. 
The lag measurements of the epoch 2020 September to 2021 June confirm the near-zero time delay between the optical and UV bands, consistent with the expectations from synchrotron theory \citep{Kirk1998}. 

\begin{figure}
\includegraphics[clip, trim=0.9cm 1.5cm 2.1cm 0.0cm, angle=-90, width=\columnwidth]{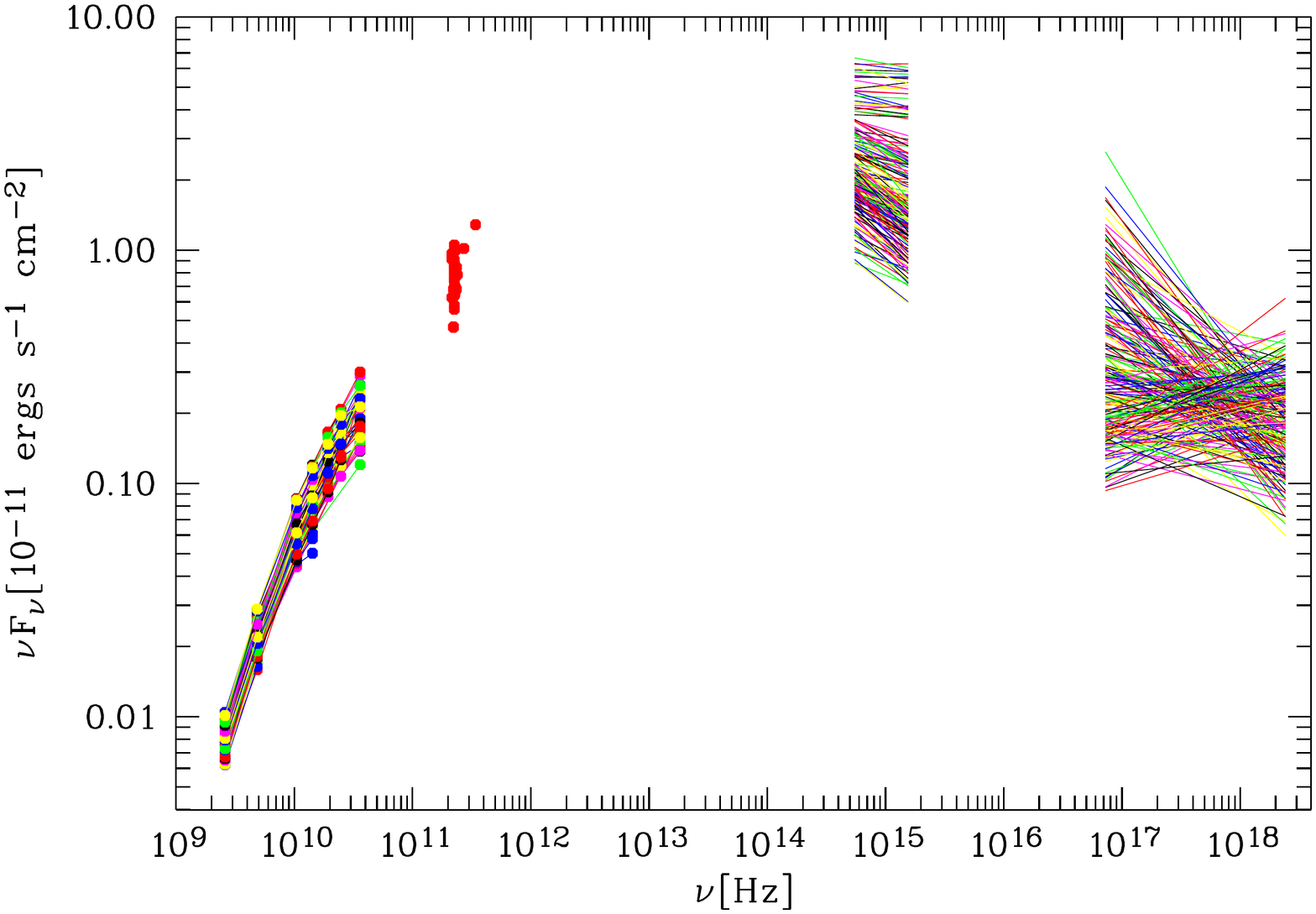}
\caption{Broadband SEDs of OJ 287 between 2019 and 2021.
Simultaneous observations in the UV--optical and in X-rays are presented by the same line colour, even though the colour scheme is meant to merely highlight the range of observed spectra, not any single one in particular. The radio observations are generally not simultaneous with the Swift data within hours, and an independent colour scheme is used. Single data points are additionally overplotted in the radio regime since not all frequencies are covered each time. 
}
\label{fig:SEDs_broad}
\includegraphics[clip, trim=0.9cm 1.5cm 2.4cm 0.0cm, angle=-90, width=\columnwidth]{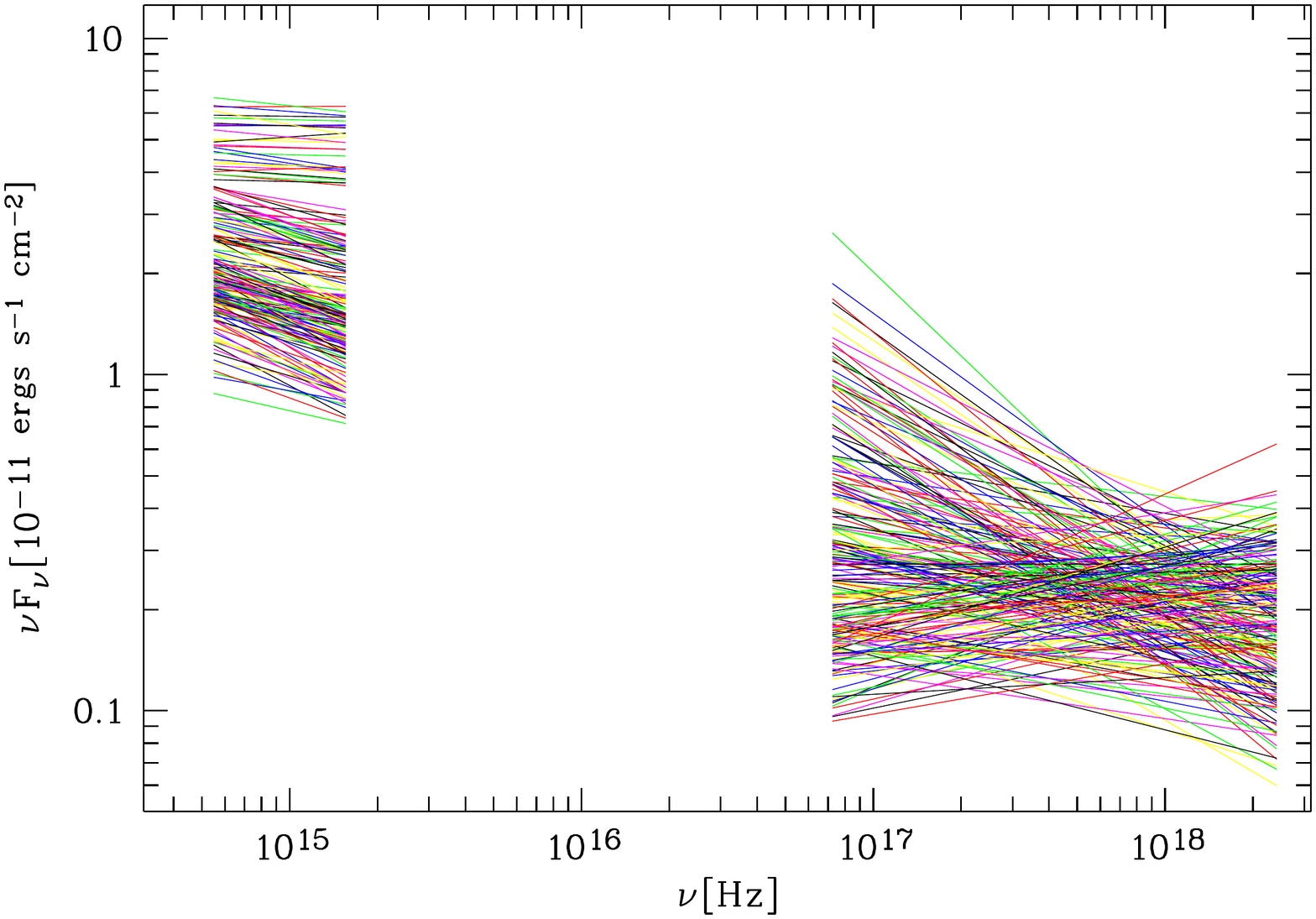}
\caption{The same SEDs as in Fig. \ref{fig:SEDs_broad}, but now a zoom on the UV--X-ray part of the SED. The trend of flatter optical--UV slopes and steeper X-ray slopes as the fluxes increase is evident.}
\label{fig:SEDs_zoom}
\end{figure}

\subsubsection{2021 December optical--UV low-state and secondary SMBH disk crossing} 

In 2021 December the UV and optical emission is at a low-state.
Such long-lasting low-flux states were last reached during the deep fade in 2017 and the 2020 September minimum \citep{Komossa2020, Komossa2021d}. 
It is interesting to note that this epoch coincides with a secondary SMBH disk impact on 2021 December 3 predicted by the SMBBH model \citep{Valtonen2022}. The related thermal impact {\em flare} will only become visible in mid-2022 according to the model predictions, once the impact-driven outflow becomes optically thin. 
However, we may expect enhanced activity due to shocks in the vein of the impact when the secondary encounters high-density material like disk winds or the disk corona prior to the actual disk impact, leading to enhanced emission in the optical to high-energy regime. 
%
Whether or not such radiation becomes detectable w.r.t. the bright long-lived non-thermal blazar emission, is another question. 
Visual inspection of the light curve in 2021 September -- December does not reveal any outstanding features.  
Further, the fractional variability analysis (Tab. \ref{tab:Fvar}) does not find an enhanced variability pattern at that epoch in comparison to the previous epoch. 

\citet{Pihajoki2013b} considered the possibility that the secondary SMBH undergoes accretion as it approaches the disk surrounding the primary SMBH, including the possibility that the secondary launches a temporary jet. 
Detailed simulations that address the question whether a classical accretion disk can form (and survive) and whether jet launching is possible under these extreme conditions, while the secondary is moving at relativistic speed through a rapidly changing environment, have not yet been carried out. If a temporary jet does form, then estimates show that it may reach a comparable radio luminosity to the primary jet under favorable conditions \citep[Sect. 8 of][]{Valtonen2022}.    
Therefore, could the strong radio flare we see rapidly rising in 2021 August be associated with the secondary SMBH ?  If so, we expect to detect the signature of the temporary accretion disk in the form of a strong X-ray signal. However, X-rays remain at low emission levels since 2021 September (when OJ 287 became observable again with Swift after the Sun constraint). Therefore, either there was only a short-lived accretion episode that had already faded in September while the jet continued to evolve and emit, or, much more likely, the radio emission we detect is from the main jet of OJ 287. One way to test further the secondary jet scenario would be to search the ongoing radio flare for rapid variability/periodicity similar to the period of the last stable orbit \citep{Pihajoki2013b, Valtaoja1985}. 

Back to the UV--optical low-state, we may ask if the secondary SMBH's disappearance itself behind the disk could contribute to the observed 2021 UV--optical inactivity directly? This is highly unlikely 
because the broad-band emission of OJ 287 is dominated by the primary SMBH within the context of the binary model, not the secondary, and therefore its disappearance behind the disk should not dim the observed broad-band emission of OJ 287. 

Another way to look at signatures of disk impact is to search for the emergence of a new spectral component during 2021 December.  
 
\begin{figure}
\includegraphics[clip, trim=0.9cm 1.5cm 2.4cm 0.0cm, angle=-90, width=\columnwidth]{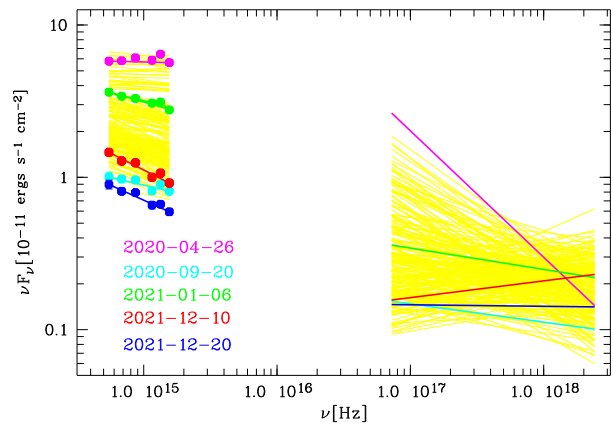}
\caption{Selected representative broadband SEDs of OJ 287 from the optical to X-rays, including high-state, low-state, and states close to candidate binary events. The yellow area represents the entirety of SEDs observed since 2019. 
}
\label{fig:SEDs_selected}
\end{figure}

If a new spectral component and/or a different emission mechanism was prevalent during this epoch, we can expect a change in the UV--optical spectral shape.
The optical--UV spectral index
$\alpha{\nu_{\rm{V-W2}}}$
of OJ 287 is in the range --1.3 to --1.5 during the month of 2021 December. 
It does not systematically change across the low-state (Fig. \ref{fig:SEDs_zoom}, \ref{fig:SEDs_selected}), and it is similar to other epochs at comparable flux low-states.  
We conclude that the low-state is driven by jet physics since there is no evidence that it could have been directly caused by the disk passage of the secondary SMBH in the binary model.    

Independent of the mechanism that causes deep minimum states, these states facilitate imaging and photometry of the host galaxy of OJ 287 \citep{Nilsson2020, Valtonen2022b} that is otherwise challenging because of the bright emission of the blazar component. In the future, the James Webb Space Telescope \citep{Gardner2006} will be ideally suited to carry out deep host imaging of OJ 287 once a new UV--optical deep fade is detected. 

The presence of the deep UV--optical low-state at a time of a high level of activity in the radio band may imply the emergence of a new or additional radio component of different broad-band emission properties. The monitoring of OJ 287 continues to follow the multi-band flux evolution. 

\section{Summary and conclusions}
We reported results from our long-term monitoring project of OJ 287, MOMO, based on 
observations taken at $>$14 different frequencies from 2.6 GHz to 100 GeV between 2019 and 2022 January, providing exquisite multiwavelength coverage at a cadence as high as one day. MOMO is a long-term project aimed at understanding disk-jet physics as well as testing new predictions of the SMBBH model of this nearby bright blazar, with the long-term goal of covering densely at least one binary orbital period of 12 yrs. 

The main results of this latest publication in a sequence can be summarized as follows: 

\begin{itemize}
\item  OJ 287 is found in a high activity phase in the radio regime with four major flare events. Turn-over frequencies are low and in the range 10--25 GHz. 
Using the sharpest, brightest flare in late 2021, an apparent brightness temperature $T_{\rm B,app}=6.7 \times 10^{12}$ K and a minimum Doppler factor $\delta_{\rm min}=3.4$ are estimated.

\item The radio--optical SED is highly variable, with strong changes in the optical/radio flux ratio of individual flare events. The well-observed non-thermal 2020 outburst \citep{Komossa2020} stands out as the one with the highest optical/radio peak flux ratio, while the major radio flare peaking in November 2021 only has a faint optical counterpart, if any.   

\item{Unlike the pre-2017 observations that were characterized by several bright, multi-week $\gamma$-ray flares, the activity in the $\gamma$-ray band has been relatively low post-2017. The two noteworthy events are (1) a $\gamma$-ray flare that is simultaneous within a day with the Spitzer IR flare  \citep{Laine2020} of 2019 July 31 and (2) the brightest $\gamma$-ray flare since 2015 that was recorded in early October 2021. } 

\item Two structures of special interest have been identified in the 2020--2021 light curves:  First, a remarkable, long-lasting "broad flare" in the UV--optical and radio bands with indications of an emerging 
reversal of the previous $\Gamma_{\rm x}$--CR correlation. The reversal can be understood if IC emission makes an increased contribution to X-rays. 
The second epoch of interest is a deep  optical--X-ray low state in 2021 December at the time of a disk crossing event predicted by the SMBBH model. 

\end{itemize}
The project MOMO continues to follow the flux and spectral evolution of OJ 287 in coming years.

\section*{Acknowledgements}
It is our pleasure to thank the Swift team for carrying out the observations we proposed and for very useful discussions. 
We would like to thank Phil Evans for very useful discussions. 
This research has made use of the
 XRT Data Analysis Software (XRTDAS) developed under the responsibility
of the ASI Science Data Center (SSDC), Italy.
This work is partly based on data obtained with the 100-m telescope of the Max-Planck-Institut
f\"ur Radioastronomie at Effelsberg.
The Submillimeter Array near the summit of Maunakea is a joint project between the Smithsonian Astrophysical Observatory and the Academia Sinica Institute of Astronomy and Astrophysics and is funded by the Smithsonian Institution and the Academia Sinica. 
The authors recognize and acknowledge the very significant cultural role and reverence that the summit
of Maunakea has always had within the indigenous Hawaiian community. We are most fortunate to have the opportunity to conduct
observations from this mountain. 
This work made use of data supplied by the UK Swift Science Data Centre at the University of Leicester.
This work has made use of Fermi-LAT data supplied by
Kocevski et al. 2021, \url{https://fermi.gsfc.nasa.gov/ssc/data/access/lat/LightCurveRepository/}.
This research has made use of the NASA/IPAC Extragalactic Database (NED) which is operated by the Jet Propulsion Laboratory, California Institute of Technology, under contract with the National Aeronautics and Space Administration.

\section*{Data Availability Statement}
Reduced data are available on reasonable request. The raw data of our project are available in the Swift data archive at \url{https://swift.gsfc.nasa.gov/archive/}.








\appendix

\section{Additional light curves} 

\begin{figure*}
\includegraphics[clip,width=15.5cm,trim=0.8cm 5.6cm 1.3cm 2.6cm]{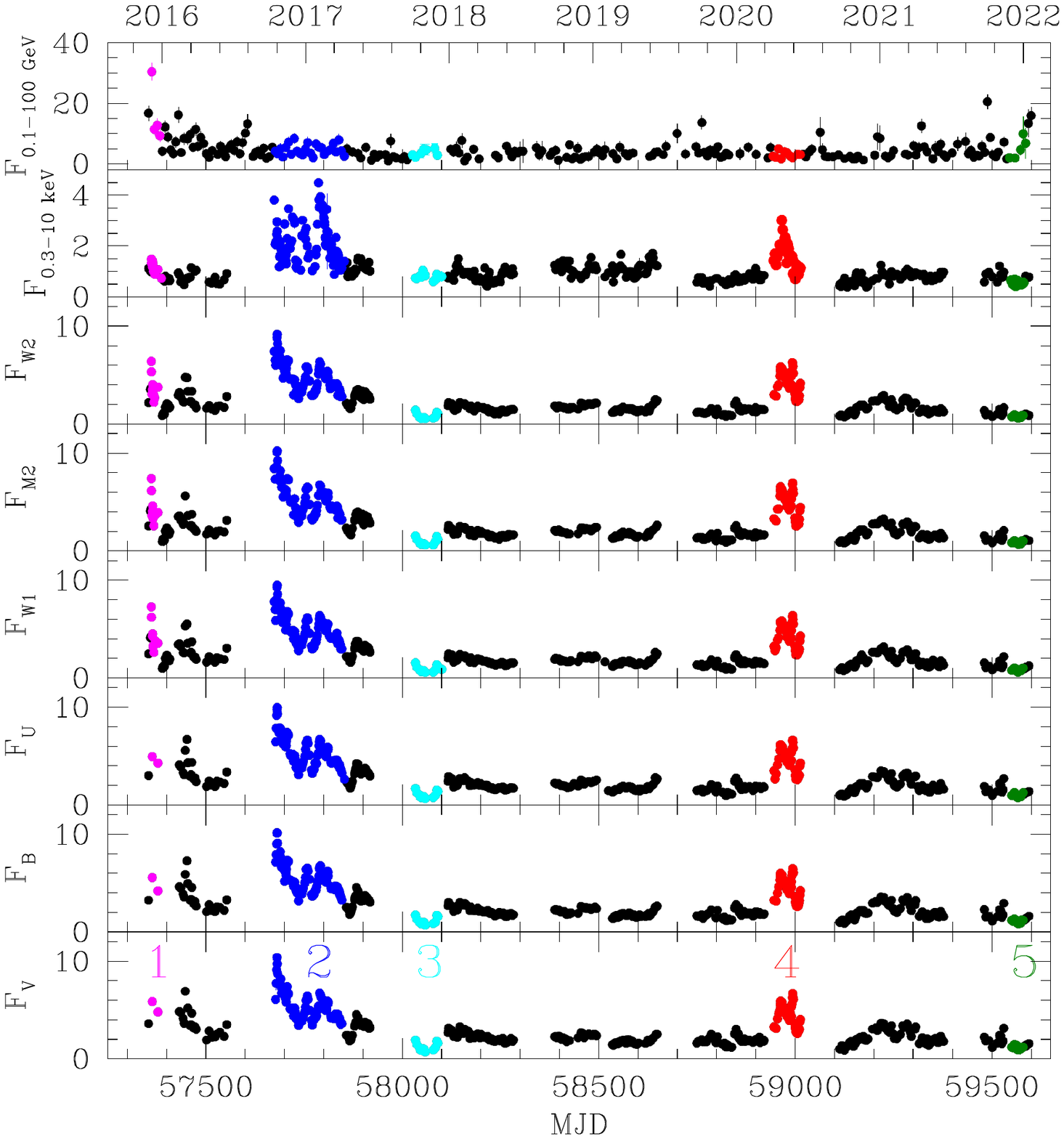}
    \caption{Fermi $\gamma$-ray and Swift X-ray to optical light curve of OJ 287 from 2015 December 1 to 2022 January 15.
    For previous versions of this light curve see \citet[e.g.,][]{Komossa2017, Komossa2020}. 
    Panels, from top to bottom: (1) $\gamma$-ray flux between 0.1-100 GeV in units of 10$^{-11}$ erg s$^{-1}$ cm$^{-2}$. (2) Absorption-corrected 0.3-10 keV X-ray flux in units of 10$^{-11}$ erg s$^{-1}$ cm$^{-2}$. (3-8) Extinction-corrected UVOT UV and optical fluxes at each filter's central wavelength in units of 10$^{-11}$ erg s$^{-1}$ cm$^{-2}$. 
    Error bars are always plotted but are often smaller than the symbol size. 
    Selected epochs are marked in color: (1) The epoch of the optical high-state in December 2015 interpreted as an impact flare in the binary model \citep[][pink]{Valtonen2016}. Note that it is associated with a $\gamma$-ray flare. (2) The epoch of the bright 2016-2017 synchrotron outburst \citep[][dark blue]{Komossa2020}. 
    (3) The epoch of the remarkable, symmetric UV--optical deep fade \citep[][light blue]{Komossa2021d}. (4) The epoch of the 2020 synchrotron outburst \citep[][red]{Komossa2020, Komossa2021a} consistent with an after-flare predicted by the binary model. 
    (5) The epoch of the UV--optical low-state in 2021 December (Sect. 6.3.4; green) that coincides in time with the binary model prediction of a disk-crossing of the secondary SMBH \citep{Valtonen2022}.   
  }
\label{fig:lc-Swift-signature}
\end{figure*}

Fig. \ref{fig:lc-Swift-signature} displays the 2015 December to 2021 December light curve of OJ 287 from the MOMO project, showing the most recent measurements in the long-term context, 
and including the Fermi data for the whole epoch. 
Particular events are marked in color including the two major outbursts in 2016/2017 and 2020, the deep low-states in 2017 and 2021, and the episode of enhanced optical--UV activity in 2015 December.     
The latter is associated with a Fermi flare
that coincides in time (within 1 day) with the sharp, bright optical flare reported by \citet{Valtonen2016}
[note that the optical peak was highest in a ground-based observation \citep{Valtonen2016} and was not caught exactly at that maximum with Swift; Fig. \ref{fig:lc-Swift-signature}]. 

\begin{figure*}
\includegraphics[clip,width=15cm,trim=0.8cm 5.6cm 0.5cm 3.0cm]{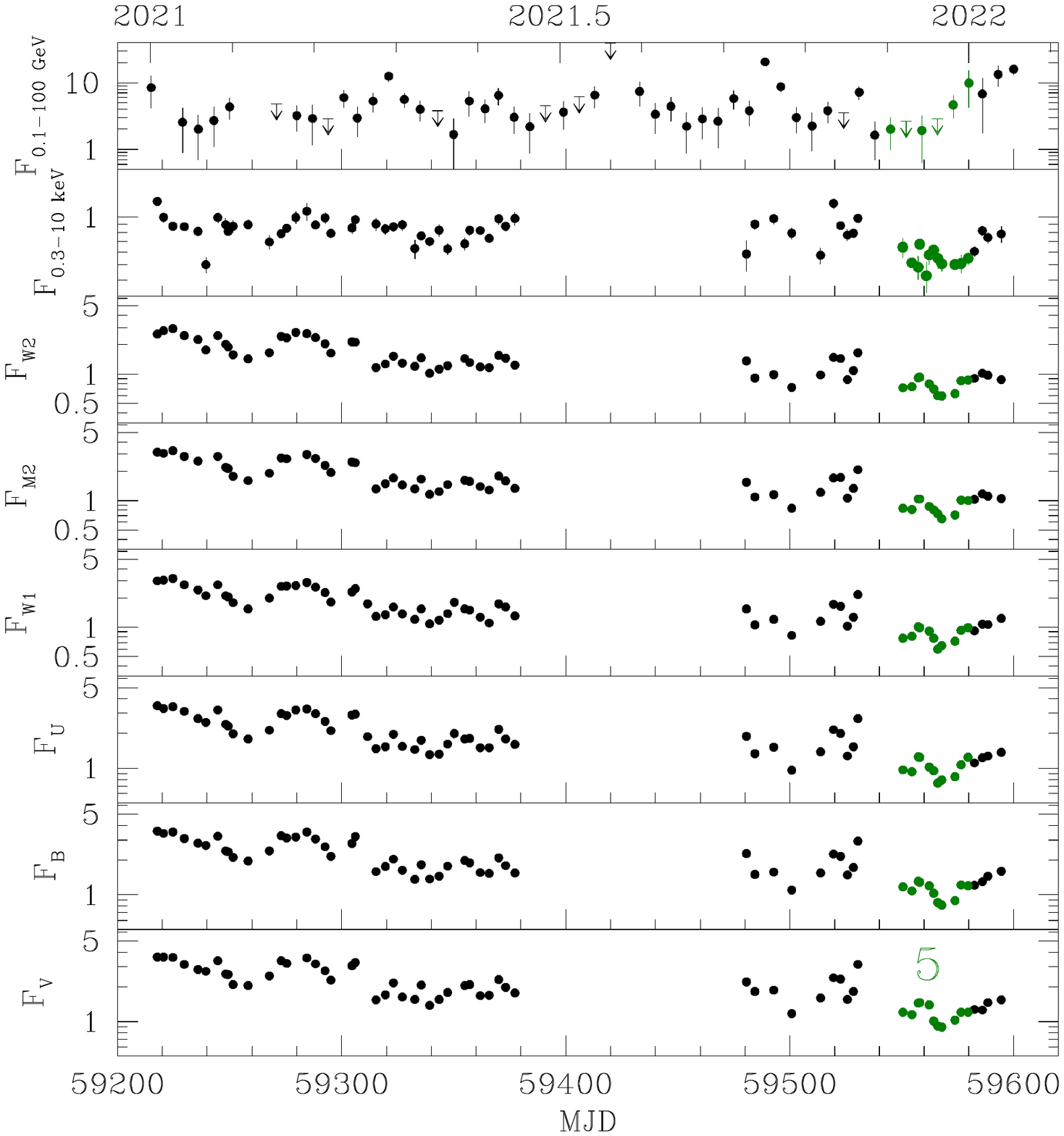}
    \caption{2021--2022.04 Swift and Fermi light curve of OJ 287. Units as in Fig. \ref{fig:lc-Swift-signature}.
    }
\label{fig:lc-Swift-2021}
\end{figure*}

In Fig. \ref{fig:lc-Swift-2021}, the Swift and Fermi observations of OJ 287 during the year 2021 are highlighted, resolving the flux evolution during this epoch in better detail than the long-term light curves. 

\bsp	
\label{lastpage}
\end{document}